\documentclass[a4paper]{JHEP3}
\usepackage[centertags]{amsmath}
\usepackage{amssymb}
\usepackage{graphicx}

\newcommand{\brk}[1]{\left [ #1 \right ]}

\newcommand{\e}{\mathrm{e}}
\newcommand{\rH}{r_{_H}}

\title{Conformal Nonlinear Fluid Dynamics from Gravity in Arbitrary
Dimensions}

\author{Sayantani Bhattacharyya$^a$\footnote{sayanta@theory.tifr.res.in}, \ R. Loganayagam$^a$\footnote{nayagam@theory.tifr.res.in}, \ Ipsita Mandal$^b$\footnote{ipsita@hri.res.in}, \ Shiraz Minwalla$^a$\footnote{minwalla@theory.tifr.res.in}, \ Ankit Sharma$^c$\footnote{ankit.iitk1@gmail.com}\\
\small{\emph{$^{a}$Department of Theoretical Physics,Tata Institute of Fundamental Research,}} \\
\small{\emph{Homi Bhabha Rd, Mumbai 400005, India.}} \\ 
\small{\emph{$^{b}$  Harish-Chandra Research Institute, Chhatnag Road, Jhusi, Allahabad 211019, India.}} \\
\small{\emph{$^{c}$ Indian Institute of Technology Kanpur, Kanpur 208016, India.}}
}

\abstract{
We generalize recent work to construct a map from the conformal Navier Stokes 
equations with holographically determined transport coefficients, 
in $d$ spacetime dimensions, to the set of 
asymptotically locally AdS$_{d+1}$ long wavelength solutions of Einstein's equations 
with a negative cosmological constant, for all $d>2$. We find simple 
explicit expressions for the stress tensor (slightly generalizing the recent
result by Haack and Yarom (\arXivid{0806.4602}) ), the full dual bulk metric and an entropy current of this strongly coupled conformal fluid, to second order in the derivative 
expansion, for arbitrary $d>2$. We also rewrite the well known exact solutions 
for rotating black holes in AdS$_{d+1}$ space in a manifestly fluid 
dynamical form, generalizing earlier work in $d=4$. To second order in the 
derivative expansion, this metric agrees with our general construction of 
the metric dual to fluid flows.
}

\keywords{}
\preprint{TIFR/TH/08-38}

\begin{document}

%*****************************************
\section{Introduction}
%*****************************************

The AdS/CFT correspondence establishes a deep connection between 
quantum field theories and theories of gravity. At generic values of 
parameters both sides of this equivalence are complicated quantum 
theories. However, every known example of the AdS/CFT duality admits 
a large $N$ limit in which the gravitational theory turns classical, 
and a simultaneous strong `t Hooft coupling limit that suppresses 
$\alpha'$ corrections to gravitational dynamics. In this limit the AdS/CFT 
correspondence asserts the equivalence between the effectively classical 
large $N$ dynamics of the local single trace operators 
$\rho_n= N^{-1}\text{Tr}\mathbf{O_n}$ of gauge theory and the classical 
two derivative equations of Einstein gravity interacting with other fields. 

The usual rules of the AdS/CFT correspondence establish a one to one map
between the bulk fields and the  single trace field theory operators; for 
instance, the bulk Einstein frame graviton maps to the field theory stress 
tensor. Given a solution of the bulk equations, the evolution of any 
given trace operator $\rho_n(x^\mu)$ may be read off from the 
normalizable fall off `at infinity' of the corresponding bulk field. This 
dictionary allows us to translate the local and relatively simple 
bulk equations into unfamiliar and extremely nonlocal 
equations for the boundary trace operators $\rho_n(x)$. The equations 
for $\rho_n(x)$ are nonlocal in both space and time; indeed the data for 
the classical evolution of $\rho_n$ includes an infinite number of time 
derivatives of $\rho_n$ on an initial slice. Given the complicated and 
unfamiliar nature of these equations, it is difficult to use our knowledge
of bulk dynamics to directly gain intuition for boundary trace dynamics. 
It would clearly be useful to identify a simplifying limit 
in order to train intuition.  

Some simplification of trace dynamics is achieved by focusing on 
a universal subsector of gravitational dynamics \cite{Bhattacharyya:2008jc} .
We focus on two derivative bulk theories of gravity that admit 
AdS$_{d+1} \times M_I$ as a solution (here $M_I$ is any internal manifold 
whose character and properties will be irrelevant for the rest of this paper).
It is easy to convince oneself that every such theory admits a consistent 
truncation to the Einstein equations with a negative cosmological constant.
The only fluctuating field under this truncation is the Einstein frame 
graviton; all other bulk fieldsare simply set to their background 
AdS$_{d+1} \times M_I$ values. This observation implies the existence of
a sector of decoupled and universal dynamics of the stress tensor in the
corresponding dual field theories. The dynamics is decoupled because all
$\rho_n(x)$ other than the stress tensor may consistently be set to zero
as the stress tensor undergoes its dynamics, and this dynamics is 
universal because the evolution of the stress tensor is governed by 
the same equations of motion in each of these infinite class of strongly coupled
CFTs. 

While the universal stress tensor dynamics described above is clearly simpler
than a general evolution of $\rho_n(x)$ in the dual theory, it is still 
both complicated and nonlocal. It is useful to take a further limit; 
to focus on boundary  configurations in which the local stress tensor 
varies on a length scale that is large, at any point,  compared to a
local equilibration length scale (intuitively, `mean free path') which is set by the 
`rest frame' energy density at the same point (we will make this more precise 
below). Local field theory intuition suggests that 
boundary configurations that obey this slow variation 
condition should be locally thermalized, and 
consequently well described by the equations of boundary fluid dynamics. 
Hence, we expect the complicated nonlocal $T_{\mu\nu}$ 
dynamics to reduce to the familiar boundary Navier Stokes equations of 
fluid dynamics in this long wavelength limit. 

All the expectations spelt out above have been demonstrated  to be true  
for $d=4,5$  from a direct analysis of the Einstein equations (See \cite{Bhattacharyya:2008jc,Loganayagam:2008is,VanRaamsdonk:2008fp,
Bhattacharyya:2008xc,Dutta:2008gf,Bhattacharyya:2008ji}). This analysis has 
also been generalized to a large extent for general $d$ in a recent paper by Haack and Yarom
\cite{Haack:2008cp}. In this paper we continue and complete the analysis 
of \cite{Haack:2008cp} in explicating the connection between the Einstein equations and fluid equations in arbitrary dimensions . 

In particular, we implement the programme initiated in \cite{Bhattacharyya:2008jc} to explicitly compute the bulk metric dual to an arbitrary fluid flow (accurate to second order in a boundary derivative expansion). We verify the expressions for the second order stress tensor dual to these flows which was recently derived in \cite{Haack:2008cp}, study the causal structure of the solutions we derive, determine their event horizons at second order in the derivative expansion, and determine an entropy current for these fluid flows. Further, we compare our results to exact solutions for rotating black holes in global AdS$_{d+1}$ and find perfect match to the expected order. In the rest of this introduction we will more carefully review some of the closely related previous work on this subject in order to place the new results of this paper in its proper context\footnote{Our main aim here is to provide the appropriate background for our work rather than to review the complete expanse of the literature relating hydrodynamics to holography. However, we have included a non-exhaustive list of references pertaining to hydrodynamics in the context of holography in the References section at the end of this paper. }.

The authors of \cite{Bhattacharyya:2008jc} developed a procedure to construct a large class of asymptotically AdS$_5$ long wavelength solutions to Einstein's equations with a negative cosmological constant. The solutions in \cite{Bhattacharyya:2008jc} were worked out  order by order in a boundary derivative expansion, and were parameterized by a four velocity field $u^\mu(x^\mu)$  and a temperature field $T(x^\mu)$. These velocity and temperature fields are further constrained to obey the four dimensional generalized Navier Stokes equations $\nabla_\mu T^{\mu\nu}=0$ where the stress tensor $T^{\mu\nu}(x^\mu)$ is a local functional of the velocity and the temperature fields. The form of $T^{\mu\nu}(x^\mu)$ was explicitly determined in \cite{Bhattacharyya:2008jc} to second order in a boundary derivative expansion (Some terms in the stress tensor were independently determined by the authors of \cite{Baier:2007ix}. Especially notable in this regard are the pioneering work in \cite{Janik:2005zt,Janik:2006ft,Heller:2007qt}.) . Consequently, the construction of \cite{Bhattacharyya:2008jc}  may be thought of as an explicit map from the space of solutions of a distinguished set of Navier Stokes equations in $d=4$ to the space of long wavelength solutions of asymptotically AdS$_{d+1}$ gravity. 

The spacetimes derived in \cite{Bhattacharyya:2008jc} were subsequently 
generalized and studied in more detail. In particular, it was demonstrated in 
\cite{Bhattacharyya:2008xc} that, subject to mild assumptions, these spacetimes have regular event horizons. In the same paper, the location of this event horizon in the `radial' direction of AdS$_5$ was explicitly  determined to second order in the derivative expansion and it was found to depend locally on the fluid data at the boundary (via a natural boundary to horizon map generated by ingoing null geodesics). The authors of \cite{Bhattacharyya:2008xc} 
also constructed a local fluid dynamical `entropy current' utilizing the 
pullback of the area form on the horizon onto the boundary. The classic 
area increase theorem of general relativity was then used to demonstrate the local form of the second law of thermodynamics (i.e., the point wise non negativity of the divergence of this entropy current). On a related note, in \cite{Loganayagam:2008is}, a formalism was developed for conformal hydrodynamics which describes the long wavelength limit of a CFT. Using this manifestly Weyl-covariant formalism, many results of \cite{Bhattacharyya:2008jc} and \cite{Bhattacharyya:2008xc} could be cast into a simpler form and Weyl covariance could be used as a powerful tool in classifying the possible forms of the metric, energy momentum tensor and the entropy current.

In \cite{Bhattacharyya:2008ji}, the construction of \cite{Bhattacharyya:2008jc} was generalized 
to spacetimes that are only locally asymptotically AdS$_5$, i.e. that asymptote to 
\begin{equation}\label{asympmetric1}
ds_5^2=\frac{1}{z^2}\left[ dz^2 + ds_{3,1}^2 \right]
\end{equation}
where $ds_{3,1}^2$ is an arbitrary slowly varying boundary metric, 
at small $z$. It is expected that, under the $AdS/CFT$ correspondence, 
such solutions are a universal subsector of the solution space of 
the relevant CFTs on the Lorentzian base manifold $M_{3,1}$ with the 
metric $ds_{3,1}^2$. In agreement with this expectation  
\cite{Bhattacharyya:2008ji} demonstrated that long wavelength solutions of gravity 
with asymptotics given by \eqref{asympmetric1} were parameterized by a velocity
and temperature field on the manifold $M_{3,1}$, subject to a covariant form 
of the Navier Stokes equations. As an example of this construction, 
the authors of \cite{Bhattacharyya:2008ji} were able to rewrite the exact asymptotically global  AdS$_5$ Kerr black hole solutions in a very simply manifestly fluid dynamical form, and demonstrate that the expansion of this metric to second order in the derivative expansion is in perfect agreement with the general construction  of the metrics dual to fluid dynamics at second order \footnote{The authors of \cite{Bhattacharyya:2008ji} also considered the coupling of a slowly varying dilaton to the metric. It would be an interesting exercise to consider generalizing the results of this paper to a bulk spacetime with dilaton dynamics. However, in this paper, we will confine ourselves to the case where the dilaton is set to its background value. On a related note , we should also mention two recent papers\cite{Erdmenger:2008rm,Banerjee:2008th} which appeared while this paper was nearing completion in which the fluid gravity correspondence was extended to a class of charged black holes in AdS$_5$ with flat boundary. }.

All of the results described above were originally worked out for the special case $d=4$, but some of these results and constructions have since been further generalized. In an early paper Van Raamsdonk 
\cite{VanRaamsdonk:2008fp} generalized the construction of the full second order bulk metric to an arbitrary fluid flow on a flat boundary to $d=3$ and also computed the holographic fluid 
dynamical stress tensor to second order in boundary derivatives. Some terms in the second order stress tensor for the uncharged conformal fluid in arbitrary dimensions were calculated using different methods by \cite{Natsuume:2007tz,Natsuume:2008iy}. Further, $1/\lambda$ and $1/N_c$ corrections to some coefficients have been computed in \cite{Buchel:2008ac,Dutta:2008gf,Buchel:2008ae,Buchel:2008kd,Buchel:2008bz}.

More recently, Haack and Yarom \cite{Haack:2008cp} partially constructed the second order bulk metric to an arbitrary fluid flow in a flat $d$ dimensional boundary (for arbitrary $d$) and fully computed the dual second order fluid dynamical stress tensor for a flat boundary. In this paper, we continue the study of Haack and Yarom \cite{Haack:2008cp} to generalize all of the work on solutions of 
pure gravity duals to arbitrary fluid flows in $d=4$ dimensions (reviewed above) to arbitrary $d>2$. 

This paper is organized as follows. In section 2 below we review the Weyl 
covariant notation for fluid dynamics that we will use in the rest of this 
paper. In section 3 we briefly explain the logic of our construction of long 
wavelength bulk solutions dual to fluid dynamics. In section 4 below we 
present explicit solutions to Einstein equations to second order in the 
boundary derivative expansion. Our solutions asymptote at small 
$z$ to   
\begin{equation}\label{asympmetric2}
ds_{d+1}^2=\frac{1}{z^2}\left[dz^2+ ds_{d-1,1}^2\right]
\end{equation}
where $ds_{d-1,1}^2$ is the arbitrarily specified weakly curved metric 
on the boundary. Our solutions are parameterized by a boundary 
 $d$-velocity field $u^\mu(x)$ and a temperature
field $T(x)$ where $x^\mu$ are the boundary coordinates. These 
velocity and temperature fields are constrained to obey 
the $d$ dimensional Navier Stokes equations, $\nabla_\mu T^{\mu\nu}=0$ where 
$T^{\mu \nu}$ is a local functional of the velocity and temperature fields. 
We also present explicit expressions for the boundary stress tensor 
$T^{\mu \nu}$ dual to our solutions.  Our answer can be expressed in an 
especially simple and manifestly Weyl-covariant form 
\begin{equation}\label{enmomintro:eq}
\begin{split}
T_{\mu\nu} &= p\left(g_{\mu\nu}+d u_\mu u_\nu \right)-2\eta \sigma_{\mu\nu}\\
&-2\eta \tau_\omega \left[u^{\lambda}\mathcal{D}_{\lambda}\sigma_{\mu \nu}+\omega_{\mu}{}^{\lambda}\sigma_{\lambda \nu}+\omega_\nu{}^\lambda \sigma_{\mu\lambda} \right]\\
&+2\eta b\left[u^{\lambda}\mathcal{D}_{\lambda}\sigma_{\mu \nu}+\sigma_{\mu}{}^{\lambda}\sigma_{\lambda \nu} -\frac{\sigma_{\alpha \beta}\sigma^{\alpha \beta}}{d-1}P_{\mu \nu}+ C_{\mu\alpha\nu\beta}u^\alpha u^\beta \right]\\
\end{split}
\end{equation}
with
\begin{equation*}
\begin{split}
b\equiv \frac{d}{4\pi T}\qquad;&\qquad p=\frac{1}{16\pi G_{\text{AdS}}b^d}\qquad;\\
\eta = \frac{s}{4\pi}=\frac{1}{16\pi G_{\text{AdS}}b^{d-1}}\qquad \text{and}& \qquad \tau_{\omega} =  b \int_{1}^{\infty}\frac{y^{d-2}-1}{y(y^{d}-1)}dy 
\end{split}
\end{equation*}
where $p$ is the pressure, $T$ is the temperature, $s$ is the entropy 
density and $\eta$ is the viscosity of the fluid. $\tau_{\omega}$ 
denotes a particular second-order transport coefficient of the fluid, 
$\sigma_{\mu\nu}$ is the shear strain rate , $\omega_{\mu\nu}$ is the vorticity and 
$C_{\mu\nu\alpha\beta}$ is the Weyl Tensor of the spacetime in which the fluid lives.
These quantities are described in more detail in section 2 below (see also Appendix A
for a list of notation used in this paper). Note that our result for the stress tensor agrees with those of Haack and Yarom\cite{Haack:2008cp} when restricted to a flat boundary manifold, but also includes an additional term proportional to boundary curvature that 
vanishes in flat space. 

In section 5 below , we demonstrate that our solutions all have a regular event horizon, and find an expression for the radial location of that event horizon upto second order in the derivative  expansion. We also construct a boundary entropy current $J_S$ that is forced by the area increase theorem of general relativity to obey the equation $\nabla_\mu J_S^\mu \geq 0$. In section 6, we rewrite the exactly known rotating black hole solutions in global AdS$_{d+1}$ in a manifestly fluid dynamical form. These solutions turn out to be dual to rigid fluid flows on $S^{d-1,1}$ (see \cite{Bhattacharyya:2007vs} for earlier work). These initially complicated looking blackhole metrics admit a rewriting in a rather simple form in the fluid dynamical gauge and variables used in this paper. In appropriate co-ordinates, the general AdS-Kerr metric\footnote{Note in particular that for $d=2$, $\sigma_{\mu\nu}=\omega_{\mu\nu}=0$ and $\mathcal{S}_{\mu\nu}$ term is absent in which case 
we get the BTZ blackhole in AdS$_3$ as shown in\cite{Haack:2008cp}.} assumes the form

\begin{equation}
\begin{split}
ds^2=&-2 u_\mu dx^\mu \left( dr + r\ \mathcal{A}_\nu dx^\nu \right) + \left[ r^2 g_{\mu\nu} +u_{(\mu}\mathcal{S}_{\nu)\lambda}u^\lambda -\omega_{\mu}{}^{\lambda}\omega_{\lambda\nu}\right]dx^\mu dx^\nu\\ 
&\qquad+ \frac{r^2 u_\mu u_\nu}{b^d\text{det}\left[r\ \delta^\mu_{\nu}-\omega^\mu{}_{\nu}\right]}  dx^\mu dx^\nu\\ 
\end{split}
\end{equation}

where $\mathcal{A}_\mu$ is the fluid dynamical Weyl-connection and
$\mathcal{S}_{\mu\nu}$ is the Weyl-covariantized Schouten tensor introduced in \cite{Loganayagam:2008is}. We demonstrate that the expansion of 
these solutions to second order in the derivative expansion agrees
with our general construction of metrics dual to fluid dynamics. 
We end our paper in section 7 with a discussion of our results 
and possible generalizations. 

Note that whereas all the results of this paper pertain apply only to $d > 2$, these results here are easily extended to the $d=2$ case which turns out however to be trivial. See Appendix B for a brief discussion of the triviality of conformal fluid dynamics in $d=2$. 

\section{Manifest Weyl Covariance}

In this section we review the Weyl covariant notation we will use in the 
rest of the paper.

\subsection{Weyl covariant notation for conformal  hydrodynamics}

The Fluid-Gravity Correspondence relates hydrodynamic solutions emerging from $CFT_{d}$ to regular Solutions of Gravity in AdS$_{d+1}$. The defining degree of freedom on the hydrodynamics side are the velocity field $u^\mu(x)$ and the temperature field $\mathcal{T}(x)$ and the energy-momentum (and an entropy current, see below) are expressed as functions of these basic degrees of freedom\footnote{In this paper, we will confine our attention to fluids with no other conserved charge except energy and momentum. Fluids with other conserved charges have additional degrees of freedom involving the chemical potentials related to those charges.}. The basic equations of hydrodynamics are the conservation of 
energy-momentum supplemented by the constitutive relations which give energy-momentum tensor as a function of the basic fields of fluid dynamics.

In this work, we will in particular focus on hydrodynamics that arises out of a $d$ dimensional CFT on a weakly curved non-dynamical metric background $g_{\mu\nu}$. The
Weyl invariance of the underlying CFT imply that the equations of hydrodynamics are   covariant under the transformations
\begin{equation}\label{eqn:weylcovdef}
\begin{split}
g_{\mu\nu}&= e^{2\phi(x)}\tilde{g}_{\mu\nu} \quad , \qquad u_\mu = e^{\phi(x)}\tilde{u}_{\mu}  \quad , \qquad T= e^{-\phi(x)}\tilde{T}\\
\end{split}
\end{equation}
Quantities (like $u^\mu(x)$) that transform homogeneously under Weyl 
transformations are said to be Weyl covariant. Notice that ordinary derivatives
of Weyl covariant quantities are not themselves Weyl covariant.  
However \cite{Loganayagam:2008is} developed a `Weyl covariant derivative', 
utilizing an effective `Weyl gauge field' built out of derivatives of the 
fluid dynamical velocity field. The formulas of conformal fluid dynamics - 
and  all the spacetime dual metrics to fluid flows that we will construct 
in this paper - may all be written in terms of the Weyl covariant derivative
of \cite{Loganayagam:2008is}, making their Weyl transformation properties 
manifest. We will extensively use the Weyl covariant notation 
developed in \cite{Loganayagam:2008is} in this paper, and so pause to 
review it in detail in this subsection. 

Let us define the `gauge field'
\begin{equation}\label{eqn:Adef}
\begin{split}
\mathcal{A}_{\nu} \equiv u^\lambda\nabla_\lambda u_{\nu}-  \frac{\nabla_\lambda  u^\lambda}{d-1} u_{\nu} = \tilde{\mathcal{A}}_{\nu} + \partial_{\nu}\phi .
\end{split}
\end{equation}
As indicated by the last part of \eqref{eqn:Adef}, the gauge field 
$\mathcal{A}_\mu$ transforms like a connection under Weyl transformations. 
This, in turn allows us  to define a Weyl-covariant derivative on arbitrary 
tensors of weight $w$. In particular, the Weyl covariant derivative of 
an arbitrary tensor $Q^{\mu\ldots}_{\nu\ldots}$ of weight $w$ (i.e. one  that 
obeys $Q^{\mu\ldots}_{\nu\ldots}= 
e^{-w\phi}\widetilde{Q}^{\mu\ldots}_{\nu\ldots}$) may be defined as  
\begin{equation}\label{eqn:Ddef}
\begin{split}
\mathcal{D}_\lambda\ Q^{\mu\ldots}_{\nu\ldots} &\equiv \nabla_\lambda\ Q^{\mu\ldots}_{\nu\ldots} + w\  \mathcal{A}_{\lambda} Q^{\mu\ldots}_{\nu\ldots} \\ &+\brk{{g}_{\lambda\alpha}\mathcal{A}^{\mu} - \delta^{\mu}_{\lambda}\mathcal{A}_\alpha  - \delta^{\mu}_{\alpha}\mathcal{A}_{\lambda}} Q^{\alpha\ldots}_{\nu\ldots} + \ldots\\
&-\brk{{g}_{\lambda\nu}\mathcal{A}^{\alpha} - \delta^{\alpha}_{\lambda}\mathcal{A}_\nu  - \delta^{\alpha}_{\nu}\mathcal{A}_{\lambda}} Q^{\mu\ldots}_{\alpha\ldots} - \ldots
\end{split}
\end{equation}
It may be checked that the covariant derivative of a tensor of weight $w$ is 
also a tensor of weight $w$.

For example the Weyl-covariant gradient of velocity 
may be decomposed into symmetric and antisymmetric pieces as 
$\mathcal{D}_\mu u_\nu = \sigma_{\mu\nu}+\omega_{\mu\nu}$ 
where $\sigma_{\mu\nu}$ is the shear strain rate of the fluid and  
$\omega_{\mu\nu}$ is the vorticity tensor. This decomposition demonstrates
each of $\sigma_{\mu\nu}$ and $\omega_{\mu\nu}$ are Weyl-covariant with
weight $w=-1$ (a property that may also be - more cumbersomely - directly
verified from the definitions of these quantities in terms of ordinary
derivatives of velocity).

Recall that in a CFT, the stress tensor transforms with the weight $d+2$ upto
the anomaly corrections and is traceless apart from the conformal
anomaly. Using these conditions, the basic equation of hydrodynamics - the conservation of the stress tensor - may be recast into a manifestly Weyl-covariant form 
\begin{equation}\label{eqn:Tmunucons}
\begin{split}
\mathcal{D}_\mu T^{\mu\nu} &\equiv \nabla_\mu T^{\mu\nu} + \mathcal{A}^\nu (T^\mu{}_\mu-\mathcal{W}) =0\\
\end{split}
\end{equation} 
where $\mathcal{W}$ denotes the conformal anomaly of the underlying CFT.

Below we will have use for additional Weyl covariant curvature tensors 
that are naturally constructed out of the Weyl covariant derivative. We 
will now define these objects 
\begin{equation}\label{eqn:Rdef}
\begin{split}
[\mathcal{D}_\mu,\mathcal{D}_\nu]V_\lambda &= w\ \mathcal{F}_{\mu\nu}\ V_\lambda + \mathcal{R}_{\mu\nu\lambda}{}^\alpha\  V_\alpha\quad \quad \text{with}\\
\mathcal{F}_{\mu\nu} &= \nabla_\mu \mathcal{A}_\nu - \nabla_\nu \mathcal{A}_\mu \\
\mathcal{R}_{\mu\nu\lambda\sigma}&= R_{\mu\nu\lambda\sigma}+ \mathcal{F}_{\mu\nu} g_{\lambda\sigma} \\
&- \delta^\alpha_{[\mu}g_{\nu][\lambda}\delta^\beta_{\sigma]}\left(\nabla_\alpha \mathcal{A}_\beta + \mathcal{A}_\alpha \mathcal{A}_\beta - \frac{\mathcal{A}^2}{2} g_{\alpha\beta} \right)\\
\end{split}
\end{equation}
where $B_{[\mu\nu]}\equiv B_{\mu\nu}- B_{\nu\mu}$ indicates antisymmetrisation
\footnote{Note that the curvature tensors appearing in this paper are negative of the curvature tensors defined in \cite{Loganayagam:2008is}}. As is evident from the expressions above, the Weyl-covariantized curvature tensors do not vanish even if the fluid lives in a flat spacetime. 

We will also need a few other tensors derived from 
the Weyl-covariantized Riemann tensor including Weyl-covariantized 
Ricci tensor $\mathcal{R}_{\mu\nu}$ and Weyl-covariantized 
Ricci scalar $\mathcal{R}$\footnote{ defined by the relations  
$\mathcal{R}_{\mu\nu}= \mathcal{R}_{\mu\lambda\nu}{}^{\lambda}$ and
$\mathcal{R}= \mathcal{R}_{\lambda}{}^\lambda$ where 
$\mathcal{R}_{\mu\nu\lambda}{}^\alpha$ is the Weyl-covariantised Riemann Tensor
defined in the eqn.\eqref{eqn:Rdef}. }. In addition, the
Weyl-covariantized Schouten tensor $\mathcal{S}_{\mu\nu}$ is defined from the 
Weyl-covariantized Ricci tensor in the same way as the
ordinary Schouten tensor.\footnote{We remind the reader that the Schouten tensor
is defined by the formula$(d-2) S_{\mu\nu} \equiv R_{\mu\nu}-\frac{R\ g_{\mu\nu}}{2(d-1)} $. One way to motivate this definition is to look at the definition of the Weyl curvature tensor $C_{\mu\nu\lambda\sigma} \equiv R_{\mu\nu\lambda\sigma}+\delta^\alpha_{[\mu}g_{\nu][\lambda}\delta^\beta_{\sigma]}S_{\alpha\beta}$.} 
\begin{equation}\label{eqn:Sdef}
\begin{split}
\mathcal{S}_{\mu\nu} &\equiv \frac{1}{d-2}\left(\mathcal{R}_{\mu\nu}-\frac{\mathcal{R}g_{\mu\nu}}{2(d-1)}\right)= S_{\mu\nu}+\left(\nabla_\mu \mathcal{A}_\nu + \mathcal{A}_\mu \mathcal{A}_\nu - \frac{\mathcal{A}^2}{2} g_{\mu\nu} \right) +\frac{\mathcal{F}_{\mu\nu}}{d-2} 
\end{split}
\end{equation} 
and in turn, the Weyl curvature $C_{\mu\nu\lambda\sigma}$ is related to its Weyl-covariant counterpart $\mathcal{C}_{\mu\nu\lambda\sigma}$ via
\begin{equation}\label{eqn:Cdef}
\begin{split}
\mathcal{C}_{\mu\nu\lambda\sigma} &\equiv \mathcal{R}_{\mu\nu\lambda\sigma}+\delta^\alpha_{[\mu}g_{\nu][\lambda}\delta^\beta_{\sigma]}\mathcal{S}_{\alpha\beta}\\
&= C_{\mu\nu\lambda\sigma} + \mathcal{F}_{\mu\nu} g_{\lambda\sigma} 
\end{split}
\end{equation}
Conformal hydrodynamics is easily formulated using the tensors above along with Weyl-covariant derivatives of velocity and temperature fields. Further, we would also find it often convenient to use the following Weyl-covariant combinations
\begin{equation}\label{eqn:sigmaRdef}
\begin{split}
u^\lambda\mathcal{D}_{\lambda} {\sigma}_{\mu\nu} &\equiv P_{\mu}{}^{\alpha}P_\nu{}^{\beta}u^{\lambda}\nabla_{\lambda}{\sigma}_{\alpha\beta} + \frac{\vartheta}{d-1}{\sigma}_{\mu\nu} \\
\mathcal{R}_{\mu\nu}&\equiv R_{\mu\nu} +(d-2)\left(\nabla_\mu \mathcal{A}_\nu + \mathcal{A}_\mu \mathcal{A}_\nu -\mathcal{A}^2 g_{\mu\nu}  \right)+g_{\mu\nu}\nabla_\lambda\mathcal{A}^\lambda + \mathcal{F}_{\mu\nu}\\
\mathcal{R} &\equiv R +2(d-1)\nabla_\lambda\mathcal{A}^\lambda - (d-2)(d-1) \mathcal{A}^2 \\
\mathcal{D}^\lambda {\sigma}_{\mu\lambda} &= \left(\nabla^\lambda-(d-1)\mathcal{A}^\lambda\right){\sigma}_{\mu\lambda} \\
\mathcal{D}^\lambda {\omega}_{\mu\lambda} &= \left(\nabla^\lambda-(d-3)\mathcal{A}^\lambda\right){\omega}_{\mu\lambda} \\
\end{split}
\end{equation}

Before ending this section we will now digress to review the motivation for our definition of the `gauge field' in structural mathematical terms\cite{Loganayagam:2008is}. The mathematically non inclined reader is invited to skip directly to the next section. 

In a spacetime which is associated with a class of Weyl-equivalent metrics, one way to write down manifestly Weyl-covariant expressions is to introduce what is known as a Weyl connection\cite{1992JMP....33.2633H,Iorio:1996ad,Manvelyan:2007tk}. A torsionless connection $\nabla^{weyl}$  is called a Weyl connection  if for every metric in a class of Weyl-equivalent metrics, there exists a one form $\mathcal{A}_\mu$ such that $\nabla^{weyl}_\mu g_{\nu\lambda}= 2 \mathcal{A}_\mu g_{\nu\lambda}$. A weyl-connection enables us to define a Weyl-covariant derivative $\mathcal{D}_\mu$ acting on a tensor field with a weight\footnote{by which we mean a tensor field which transforms as $Q^{\mu}_{\nu}=e^{-w\phi}\tilde{Q}^{\mu}_{\nu}$ under Weyl transformation.} $w$  as  $\mathcal{D}_\mu\equiv \nabla^{weyl}_\mu + w\mathcal{A}_\mu $. In terms of $\mathcal{D}_\mu$, the above requirement on the connection becomes $\mathcal{D}_{\lambda}g_{\mu\nu} =0$ which is just the statement of metric compatibility of the Weyl-covariant derivative. Such a mathematical structure is especially relevant to the 
problem at hand since hydrodynamics on a spacetime background provides us with a natural Weyl connection. The required $\mathcal{A}_\mu$ is uniquely determined  by requiring that $u^\lambda\mathcal{D}_\lambda u^\mu = 0 $ and $\mathcal{D}_\lambda u^\lambda =0 $. Using these equations, we can solve for $\mathcal{A}_\mu$ in a particular Weyl frame in terms of the velocity field $u^\mu$ and the usual\footnote{defined via the Christoffel connection of the metric in that frame} covariant derivative $\nabla_\mu$  to get the equation~\eqref{eqn:Adef}.

%The hydrodynamics in the boundary theory is specified by giving the boundary energy-momentum tensor as a function of the velocity/temperature fields and their derivatives. We can expand the energy-momentum tensor of a conformal fluid in a boundary derivative expansion as
%\begin{equation}\label{eqn:Tmunuexp}
%T_{\mu\nu} = p\ \left(g_{\mu\nu}+ d u_\mu u_\nu \right) + \pi_{\mu\nu} 
%\end{equation}
%where $p$ denotes the pressure of the fluid and $\pi_{\mu\nu}$ denotes the viscoelastic stress distribution in a fluid. This viscoelastic stress distribution specifies the various internal forces acting on the fluid element with a given velocity/temperature gradient. In conformal hydrodynamics, the energy-momentum tensor is traceless (except for the conformal anomaly) - this dictates the above form of the energy-momentum tensor.
%We shall follow the convention of Landau and Lifshitz\cite{1959flme.book.....L} for the velocity field $u^\mu$, i.e., we take $u^\mu$ to be the velocity of energy transport. Together these imply that $\pi_{\mu\nu}$ is both transverse ($u^\mu\pi_{\mu\nu}=0$) and traceless ($\pi^{\mu}{}_\mu=0$) except for conformal anomaly.

\subsection{Classification of fluid dynamical Weyl covariant Tensors to second order}

In this subsection, we will classify all the Weyl invariant scalars,
transverse vectors (a vector/tensor is transverse if it is orthogonal 
to $u^\mu$) and  symmetric traceless transverse tensors with less than or
 equal to two derivatives. We enumerate these invariants `on shell'. In 
other words two invariants that are set equal by the equations of motion
 of fluid dynamics are treated as equal in our counting.

In order to classify the Weyl-covariant tensors, we begin with the basic quantities of hydrodynamics - the fluid temperature $T\equiv d/(4\pi b)$ and the fluid velocity $u^\mu$ . The temperature is a Weyl-covariant scalar with conformal weight unity and the velocity is a 
Weyl-covariant vector with conformal weight unity. Consequently ${b u^\mu}$
is the only derivative free conformally invariant object one can build out of
the basic fields of fluid dynamics. It follows that, at the zero derivative level,
there are no non-trivial Weyl-invariant scalars, no invariant transverse vector
or  symmetric traceless transverse tensors.

Let us now turn to one derivative Weyl invariants of fluid dynamics. Recall that we
wish to classify only the on-shell invariants. Clearly the relations imposed 
by the $\nabla_\mu T^{\mu\nu}=0$ on one derivative invariants have their origin 
in `zero derivative' or perfect fluid contributions to the stress tensor.
This tensor is proportional to $b^{-d}\left( g_{\mu\nu}+d u_\mu u_\nu \right)$.  The conservation of this stress tensor allows us to express all derivatives of $b$ in terms of derivatives of the velocity(see Appendix C). It follows that, with our 
rules of counting, at the one derivative level, there are no Weyl-invariant 
scalars or transverse vectors. However there is one Weyl-invariant symmetric 
traceless transverse tensor $b \sigma_{\mu\nu}$ at the first derivative 
level\footnote{In this counting, we do not count the pseudotensors since they never occur
in our computations.}.

In order to enumerate distinct on-shell Weyl invariants at the two derivative level, it is important to account for the contribution of one derivative terms in the stress tensor to the equation of motion. As we have explained above, the only possible one derivative correction to the stress tensor (in the Landau Lifshitz gauge, see ahead for an explanation) is proportional to $\sigma_{\mu\nu}$. After analyzing the two derivative relations that follow from the conservation of the corrected stress tensor, it is not difficult to show there are three independent Weyl-invariant scalars (which are $ b^2 \sigma_{\mu\nu} \sigma^{\mu\nu} $ , $ b^2 \omega_{\mu\nu} \omega^{\mu\nu} $ and $b^2\mathcal{R}$), two Weyl-invariant transverse vectors ( $b P_\mu^{\nu} \mathcal{D}_\lambda \sigma_{\nu}{}^\lambda$ and $b  P_\mu^{\nu} \mathcal{D}_\lambda \omega_{\nu}{}^\lambda$) and five Weyl-invariant symmetric traceless transverse tensors - 
\begin{equation}\label{eqn:2derivTens}
\begin{split}
u^\lambda\mathcal{D}_\lambda\sigma_{\mu\nu},\qquad &C_{\mu\alpha\nu\beta}u^\alpha u^\beta ,\qquad \omega_{\mu}{}^{\lambda}\sigma_{\lambda\nu}+\omega_{\nu}{}^{\lambda}\sigma_{\lambda\mu},\\
\sigma_{\mu}{}^{\lambda}\sigma_{\lambda\nu}-\frac{P_{\mu\nu}}{d-1}\  \sigma_{\alpha\beta}\sigma^{\alpha\beta}\qquad  &\text{and}\qquad
\omega_{\mu}{}^{\lambda}\omega_{\lambda\nu}+\frac{P_{\mu\nu}}{d-1}\  \omega_{\alpha\beta}\omega^{\alpha\beta} .\\
\end{split}
\end{equation}

We will find this classification of Weyl invariants very useful in the rest of 
this paper. For instance, the classification immediately implies that 
the energy momentum tensor of a general fluid configuration to second 
order in derivative expansion should assume the form 
\cite{Baier:2007ix,Loganayagam:2008is}
\begin{equation}\label{Tgeneral:eq}
\begin{split}
T^{\mu\nu}&= p (g^{\mu\nu}+d u^\mu u^\nu) \\
&\quad -2\eta \brk{\sigma^{\mu\nu}-\tau_1\ u^\lambda \mathcal{D}_\lambda \sigma^{\mu\nu}+\tau_2(\omega^\mu{}_\lambda\sigma^{\lambda\nu}+\omega^\nu{}_\lambda\sigma^{\lambda\mu}) }\\
&\quad +\xi_\sigma [\sigma^\mu{}_\lambda\sigma^{\lambda\nu}-\frac{P^{\mu\nu}}{d-1}\sigma^{\alpha\beta}\sigma_{\alpha\beta}]+\xi_C\ C_{\mu\alpha\nu\beta}u^\alpha u^\beta\\
&\quad +\xi_\omega [\omega^\mu{}_\lambda\omega^{\lambda\nu}+\frac{P^{\mu\nu}}{d-1}\omega^{\alpha\beta}\omega_{\alpha\beta}]+\ldots\\
\end{split}
\end{equation}
where Weyl-covariance demands that
\begin{equation}\label{TcoeffWeylDep:eq}
\begin{split}
p\propto b^{-d}\ ,\quad \eta\propto b^{1-d}\ , \quad \tau_{1,2}\propto b \ ,\quad \xi_{\sigma,C,\omega} \propto b^{2-d}
\end{split}
\end{equation}

\section{Perturbative Construction of Solutions}

In this section, we will briefly review the basic logic 
that underlies the construction of gravity solutions dual to 
arbitrary fluid flows. The methodology employed in this paper is 
an almost direct generalization of the techniques used in 
\cite{Bhattacharyya:2008jc,Loganayagam:2008is,VanRaamsdonk:2008fp,
Bhattacharyya:2008xc,Dutta:2008gf,Bhattacharyya:2008ji,Haack:2008cp}. 
Consequently in this paper we will describe the logic of our construction
and the details of implementation only briefly, referring the reader to 
the references above for more details.

\subsection{Equations of motion and uniform brane solutions}

In this paper we develop a systematic perturbative expansion to solve 
Einstein's equations with a negative cosmological constant
\begin{equation}\label{eqn:Eineq}
\mathcal{G}_{AB}-\frac{d(d-1)}{2} G_{AB}= 0, ~~~M, N=1 \ldots d+1
\end{equation}
where $\mathcal{G}_{AB}$ denotes the Einstein tensor of the bulk metric 
$G_{AB}$.

One solution of these equations is AdS spacetime of unit radius  
\begin{equation}\label{eqn:Ads}
ds^2=\frac{dr^2}{r^2} +r^2 \left( \eta_{\mu\nu} dx^\mu dx^\nu \right), ~~~
\mu, \nu=1 \ldots d
\end{equation}
Other well known solutions to these equations include boosted black branes 
which we write here in Schwarzschild like coordinates
\begin{equation} 
\label{eqn:Adsbrane}
\begin{split}
ds^2&=\frac{dr^2}{r^2 f(r) } +r^2 \left( -f(r) u_\mu u_\nu dx^\mu dx^\nu 
+ {\mathcal P}_{\mu\nu} dx^\mu dx^\nu \right)  \\
f(r)&= 1-\frac{1}{(br)^d}, ~~~g_{\mu \nu} u^\mu u^\nu=-1, ~~~
{\mathcal P}_{\mu\nu}=g_{\mu\nu}+u_\mu u_\nu, ~~~b=\frac{d}{4 \pi T}
\end{split}
\end{equation}
$g_{\mu\nu}$ in \eqref{eqn:Adsbrane} is an arbitrary 
constant boundary metric of signature $(d-1, 1)$, while $u^\mu$ in 
the same equation is any constant unit normalized $d$ velocity. 
Of course any constant metric of signature $(d-1,1)$ can be 
set to $\eta_{\mu\nu}$ by an appropriate linear coordinate 
transformation $x^\mu \rightarrow \Lambda^\mu_\nu x^\nu$,
and $u^\mu$ can subsequently be set to $(1, 0 \ldots 0)$ by a boundary Lorentz 
transformation. Further $b$ in \eqref{eqn:Adsbrane} may also be set to 
unity by a coordinate change; a uniform rescaling of boundary coordinates 
coupled with a rescaling of $r$. Thus the $d(d+3)/2$ parameter set of metrics 
\eqref{eqn:Adsbrane} are all coordinate equivalent. Nonetheless we will 
find the general form \eqref{eqn:Adsbrane} useful below; indeed we will 
find it useful to write \eqref{eqn:Adsbrane} in an even more general 
coordinate redundant form. Consider 
\begin{equation} 
\label{eqn:Adsbranen}
ds^2=\frac{ \left( d\tilde{r} + \tilde{r}  \mathcal A_\nu dx^\nu \right)^2 
}{\tilde{r}^2 f(r) } +\tilde{r}^2 \left( -f(\tilde{b}\tilde{r}) 
\tilde{u}_\mu \tilde{u}_\nu dx^\mu dx^\nu 
+ {\tilde{\mathcal P}}_{\mu\nu} dx^\mu dx^\nu \right) 
\end{equation}
where 
\begin{equation}\label{eqn:weyldef}
\tilde{g}_{\mu\nu}=\e^{2\phi(x^\mu)} g_{\mu\nu}, \quad 
\tilde{u}_\mu =\e^{\phi(x^\mu)} u_\mu, \quad \tilde{b} = 
\e^{\phi(x^\mu)} b, 
\end{equation}
$\phi(x^\mu)$ is an arbitrary function and $g_{\mu\nu}$, $u^\mu$, and 
$b$ are as defined in the previous equation. This metric is coordinate
 equivalent to \eqref{eqn:Adsbrane} under the 
variable transformation $\tilde{r}\mapsto \e^{-\phi} r$. Consequently, 
the whole function worth of spacetimes \eqref{eqn:Adsbranen} (taken 
together with the restrictions \eqref{eqn:weyldef}) are all exact 
solutions to Einstein's equations and are all coordinate equivalent.

While the metrics \eqref{eqn:Adsbranen} all describe the same bulk geometry, 
in this paper we will give these spacetimes distinct  though Weyl equivalent
boundary interpretations by regulating them inequivalently near the boundary. 
We will choose to regulate the spacetimes \eqref{eqn:Adsbranen} on 
slices of constant ${\tilde r}$ and consequently regard them as states 
in a conformal field theory on a space with metric ${\tilde g}_{\mu\nu}(x)$.
With this convention, (the non-anomalous part of) the boundary stress tensor dual to \eqref{eqn:Adsbranen} is given by 
\begin{equation}
T_{\mu\nu} = \frac{1}{16\pi G_{\text{AdS}} \tilde{b}^d}
\left(\tilde{g}_{\mu\nu}+d \tilde{u}_\mu \tilde{u}_\nu \right)
\end{equation}
which shows that the metric in the Equation\eqref{eqn:Adsbranen} 
is dual to a conformal fluid with a pressure $p=1/(16\pi G_{\text{AdS}} b^d)$ 
and without any vorticity or shear strain rate. Of course the boundary 
configurations dual to \eqref{eqn:Adsbranen} with equal
$g_{\mu\nu}, u^\mu, b$ but different values of $\phi$ are related to 
each other by boundary Weyl transformations.

Notice that 
\begin{equation} \label{eqn:branest}
T^\mu_\nu u^\nu = \frac{K}{b^d} u^\mu, ~~~K=-\frac{(d-1)}{16 \pi 
G_{\text{AdS}}}
\end{equation}
In other words the velocity field is the unique time like eigenvector 
of the stress tensor, and the inverse temperature field $b$ is simply 
related to its eigenvalue. We will use this observation in the next 
subsection.

\subsection{Slow variation and bulk tubes and our zero order ansatz}

Consider an arbitrary locally asymptotically AdS$_{d+1}$ solution to 
Einstein's equations \eqref{eqn:Eineq} whose dual boundary stress tensor 
everywhere has a unique timelike eigenvector. Let this eigenvector 
(after unit normalization) be denoted by $u^\mu(x)$ and the corresponding 
eigenvalue by $\frac{K}{b^d}$. We define $u^\mu(x)$ to be the $d$ velocity 
field dual to our solution, and also define $b(x)$ to be the inverse 
temperature field dual to our solution. 

Let $\delta x(y)$ denote smallest length scale of variation of the stress 
tensor of the corresponding solution at the point $y$. We say that the 
solution is `slowly varying' if everywhere  $\delta x (y) \gg  b(y)$. 
(As will be apparent from our final stress tensor below, $b(y)$ may be 
interpreted as the effective length scale of equilibration of the field 
theory at $y$).  Similarly, we say that the boundary metric is weakly 
curved if $b(y)^2 R(y) \ll 1$ (where $R(y)$ is the curvature scalar, or 
more generally an estimate of the largest curvature scale in the problem). 

In the previous section we described uniform brane solutions of Einstein's 
equations. In the appropriate Weyl frame the temperature, velocity, boundary
metric and hence the stress tensor of those configurations was constant 
in boundary spacetime. These configurations are exact solutions to 
Einstein's equations. We will now search for solutions to Einstein equations 
with slowly varying (rather than constant) boundary stress tensors on 
a boundary manifold that has a weakly curved (rather than flat) boundary 
metric. 
From field theory intuition we expect all such boundary configurations 
to be locally patchwise equilibriated (but with varying values of the boundary 
temperature and velocity fields). This suggests that the corresponding 
bulk solutions should approximately be given by patching together 
tubes of the uniform black brane solutions. We expect these tubes 
to start along local patches on the boundary and then extend into the 
bulk following an ingoing `radial' curve. However this 
expectation leaves open an important question: what is the precise shape 
of the radial curves that our tubes follow?

One guess might be that the tubes follow the lines 
$x^\mu=$constant in the Schwarzschild coordinates we have employed so far. 
According to this guess, the bulk metric dual to slowly varying 
boundary stress tensors and boundary metric is approximately given by 
\begin{equation} 
\label{eqn:Adsbranenv}
ds^2=\frac{\left( d\tilde{r} + \tilde{r} \mathcal A_\nu dx^\nu \right)^2}
{\tilde{r}^2 f(r) } +\tilde{r}^2 \left( -f(\tilde{b}\tilde{r}) 
\tilde{u}_\mu \tilde{u}_\nu dx^\mu dx^\nu 
+ {\tilde{\mathcal P}}_{\mu\nu} dx^\mu dx^\nu \right) 
\end{equation}
where 
\begin{equation}\label{eqn:weyldefn}
\tilde{g}_{\mu\nu}=\e^{2\phi} g_{\mu\nu}, \quad \tilde{u}_\mu=
\e^{\phi} u_\mu, \quad \tilde{b} = \e^{\phi} b, 
\end{equation}
where $g_{\mu\nu}(x)$ is a weakly curved boundary metric, and $u_\mu(x)$ 
and $b(x)$ are slowly varying boundary functions. 

Although this guess seems natural, we believe it is wrong. The technical 
problem with this guess is that the metric of \eqref{eqn:Adsbranenv} does 
not in general have a regular future 
horizon\cite{Reall:private} (for particular examples of similar metrics 
that do not have a regular future horizon see \cite{Chamblin:1999by,Benincasa:2007tp,Buchel:2008kd}.The last two references show a boost-invariant expansion that develops a singular future horizon). In this paper
we will be interested only in regular solutions of Einstein equations; 
solutions whose (future) singularities are all shielded from the boundary 
of AdS by regular event horizons. As any perturbation to \eqref{eqn:Adsbranenv} that turns it into a regular space must necessarily be 
large in the appropriate sense, it follows that \eqref{eqn:Adsbranenv} 
is not a good starting point for a perturbative expansion of the solutions 
we wish to find. 

There is another more intuitive problem with the proposal that 
the ansatz \eqref{eqn:Adsbranenv} is dual to boundary fluid dynamics.
It is an obvious fact about fluid dynamical evolution that the initial 
conditions of a fluid may be chosen independent of any `kick' (forcing)
one may choose to apply to the fluid at a later time.   It seems reasonable 
to expect the same property of the bulk solutions dual to fluid dynamics. 
\footnote{In our set up we can
kick our fluid at $y^\mu$ by varying the boundary metric at $y^\mu$ (this 
induces an effective force on the fluid).} Now consider kicking a fluid in 
an arbitrary motion at the point $y^\mu$. The future evolution
of the fluid is affected only in the `fluid causal future' - of $y^\mu$. 
We call this region $C(y^\mu)$. Note that $C(y^\mu)$ lies within the future boundary 
light cone of $y^\mu$ \footnote{This is strictly true only if we sum all orders in 
the fluid expansion. Truncation at any finite order could lead to apparent violations 
of causality over length scales of order $1/T$.}. Now consider the bulk region 
$B(y(\mu)$ that consists of the union of all the tubes, referred to above, that originate
in the boundary region $C(y^\mu)$. Clearly $B(y^\mu)$ is the part of the bulk spacetime
that is affected by our kick at $y^\mu$. Bulk causality implies that $B(y^\mu)$ must lie
entirely within the future bulk light cone of $y^\mu$. 

This requirement is not met if we generate $B(y^\mu)$ with our tubes that run along lines of constant $x^\mu$ in Schwarzschild coordinates. However it is met in a particularly 
natural way (given the massless nature of the graviton) if our tubes are chosen to run along ingoing null 
geodesics. \footnote{For a related discussion on the desirability of using ingoing null geodesic tubes vis a vis causality violating tubes, see \cite{Heller:2008mb,Kinoshita:2008dq}.}

% Figure 
\begin{figure}[ht!]
 \begin{center}
\includegraphics[scale=0.8]{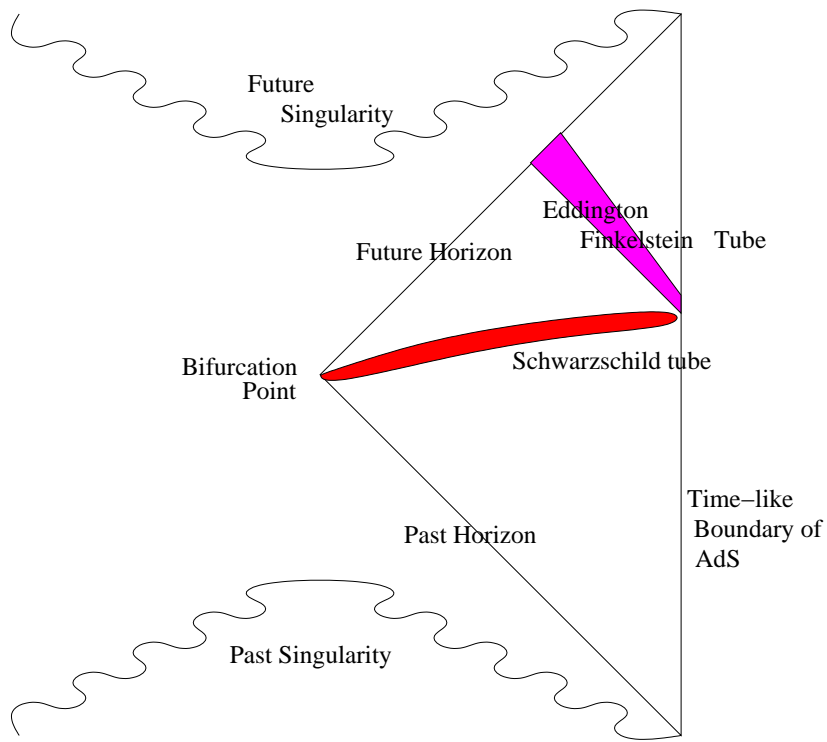} 
\end{center}
\caption{Penrose diagram of the uniform black brane illustrating the causal Eddington-Finkelstein tubes running along ingoing null geodesics . The tubes
 with $x^\mu_{\text{Schwarzschild}}$=constant are also shown. Note that we have
suppressed the other regions of the penrose diagram not germane to the discussion
in this paper. }
\label{AdSSch}
\end{figure}

With this discussion in mind, let us consider the ansatz
\begin{equation}\label{eqn:weylboostbrane}
\begin{split}
ds^2 &= -2 u_\mu dx^\mu \left( dr + r\ \mathcal{A}_\nu dx^\nu \right) + r^2 g_{\mu\nu} dx^\mu dx^\nu+\frac{r^2}{(br)^d} u_\mu u_\nu dx^\mu dx^\nu
\end{split}
\end{equation}
where once again 
\begin{equation}\label{eqn:weyldefnn}
\tilde{g}_{\mu\nu}=\e^{2\phi} g_{\mu\nu}(x), \quad \tilde{u}_\mu= 
\e^{\phi} u_\mu(x), \quad \tilde{b} = \e^{\phi} b(x), 
\end{equation}
and $g_{\mu\nu}(x)$ is a weakly curved boundary metric, and $u_\mu(x)$ 
and $b(x)$ are slowly varying boundary functions. When $g_{\mu\nu}$ $u_\mu$ 
and $b$ are all constant, \eqref{eqn:weylboostbrane} is once again simply the uniform brane solution, rewritten in Eddington Finkelstein coordinates; i.e. 
when  $g_{\mu\nu}$ $u_\mu$ and $b$ are all constant 
\eqref{eqn:weylboostbrane} and \eqref{eqn:Adsbranen} are coordinate 
equivalent(via large co-ordinate transformations). 

However when $g_{\mu\nu}$ $u_\mu$ and $b$ are functions
of $x^\mu$ \eqref{eqn:weylboostbrane} and \eqref{eqn:Adsbranenv} are 
inequivalent and in fact differ  qualitatively. As we will demonstrate below, 
under mild assumptions the metric in \eqref{eqn:weylboostbrane} has a regular event horizon that shields all the boundary from all future singularities in this space. Consequently, 
this space may (unlike the spacetime in \eqref{eqn:Adsbranen}) legitimately 
be used as the first term in the perturbative expansion of a regular 
solution of Einstein's equations. Moreover the space described in 
\eqref{eqn:weylboostbrane} approximates the uniform 
brane solution along tubes of constant $x^\mu$ in \eqref{eqn:weylboostbrane}; 
such tubes approximately follow null ingoing geodesics in this space. 

For all these reasons, in the rest of this paper we will use 
\eqref{eqn:weylboostbrane} as the first term in a systematic perturbative
expansion of a regular solution to Einstein's equations. The perturbative 
expansion parameter is $\frac{1}{b \delta x}$ (we assume that the curvature
scale in the metric is of the same order as $1/\delta x$). We emphasize 
that the solutions we find could not be obtained in a legitimate 
perturbation expansion, starting from \eqref{eqn:Adsbranen}. Several authors
have attempted to obtain the bulk metric dual to a `boost invariant 
Bjorken fluid flow' starting with the zero order solution described by 
Janik and Peschanski\cite{Janik:2005zt}, and correcting it in an expansion in $1/\delta x b$ 
(that turns into an expansion in $1/\tau^{\frac{2}{3}}$ for those 
particular solutions). As pointed out in \cite{Heller:2008mb,Kinoshita:2008dq}, however, the zeroth order solution of Janik and Peschanski is precisely \eqref{eqn:Adsbranen} for 
the particular case of boost invariant flow. Consequently, while the 
approach of the current work and \cite{Bhattacharyya:2008jc, 
VanRaamsdonk:2008fp, Dutta:2008gf, Bhattacharyya:2008ji, Haack:2008cp}
are similar in spirit to the perturbation procedure initiated by Janik and 
Peschanski, we differ at a crucial point. While those authors effectively 
adopt \eqref{eqn:Adsbranen} as the starting point of their perturbation 
theory (for the single solution they consider), in our work we adopt 
the inequivalent and qualitatively different space \eqref{eqn:weylboostbrane}
as the starting point of our perturbative expansion. 
 
\subsection{Perturbation theory at long wavelengths}

The logic behind - and the method of implementation of - this perturbative procedure have 
been described in detail in \cite{Bhattacharyya:2008jc} and also 
in \cite{VanRaamsdonk:2008fp, Dutta:2008gf, Bhattacharyya:2008ji, 
Haack:2008cp}. It has also been described in those papers how this perturbative
procedure establishes a map between solutions of fluid dynamics and 
regular long wavelength solutions of Einstein gravity with a negative 
cosmological constant. The discussion in the cited references applies almost
without modification to the current work, so we describe it only very briefly.  

We start with the ansatz $g_{MN}=g^{(0)}_{MN}+\epsilon g^{(1)}_{MN}+ \epsilon^2 
g^{(2)}_{MN} + \ldots$. Here $g^{(0)}_{MN}$ is given by \eqref{eqn:weylboostbrane}, 
$\epsilon$ is the small parameter of the derivative expansion, and 
$g^{(k)}_{MN}$ are the corrections to the bulk metric that we will determine 
with the aid of the bulk Einstein equation. 

In implementing our perturbative procedure we adopt a choice of gauge. 
As in all the metrics described above, we use the coordinates $r, x^\mu$ 
for our bulk spaces. We use $x^\mu$ as coordinates that parameterize 
the boundary and $r$ is a radial coordinate. In order to give precise 
meaning to our coordinates we need to adopt a choice of gauge. In this paper 
we choose the gauge $g_{rr}=0$ together with $g_{r\mu}=-u_\mu$. 
The geometrical implication of this gauge choice was discussed in \cite{Bhattacharyya:2008ji}, where it was explained that with this
choice lines of constant $x^\mu$ are ingoing null geodesics along 
each of which $r$ is an affine parameter. Note that the gauge choice 
described in this paper is different in detail 
from that employed in \cite{Bhattacharyya:2008jc} and also 
in \cite{VanRaamsdonk:2008fp, Dutta:2008gf, Bhattacharyya:2008ji, 
Haack:2008cp}.

The Bulk Einstein equations decompose into `constraints' on the boundary hydrodynamic data and `dynamical equations' for the bulk metric  along the tubes which are solved order by order in the derivative expansion. The dynamical equations determine the corrections that should be added to our initial metric to make it a solution of the Einstein equations. At each order, we get inhomogeneous linear equations -but, with the same homogeneous parts. These inhomogeneous linear equations obtained from Einstein equations can be solved order by order by imposing regularity at the zeroth order future horizon and appropriate asymptotic fall off at the boundary.These boundary conditions - together with a clear definition of velocity, which fixes the ambiguity of adding zero modes - give a unique 
solution for the metric, as a function of the original boundary velocity and temperature 
profile inputted into the metric $g^{(0)}_{MN}$ - order by order  in the boundary derivative expansion.

Now, we turn to the `constraints'. The `constraints' on the boundary data can be shown to be equivalent to the requirement of the conservation of the boundary stress tensor. Recall 
that we have already used the dynamical Einstein equations to determine the full bulk 
metric - and hence the boundary stress tensor - as a function of the input velocity 
and temperature fields. It follows that the constraint Einstein equations reduce simply 
to the equations of fluid dynamics, i.e. the requirement of a conserved stress tensor 
which, in turn, is a given function of temperature and velocity fields. 

It may be worthwhile to reiterate that,  as expected from fluid-gravity correspondence, metric duals which solve Einstein equations can be constructed only for those fluid configurations which solve the hydrodynamic equations. In the next section, we will present the metric which is obtained by adopting this procedure.

\subsection{Weyl Covariance}

In this subsection we explain that the bulk metrics dual to fluid dynamics 
must transform covariantly under boundary `Weyl' transformations. 
See  \cite{Bhattacharyya:2008ji} for a more detailed explanation of this fact.  

To start with we note that our bulk gauge choice (described in the previous subsection) is Weyl covariant. Any metric that obeys that gauge choice can be put in the form  
\begin{equation}\label{eq:metstd}
\begin{split}
ds^2 &= -2 u_\mu(x) dx^\mu(dr+\mathcal{V}_\nu(r,x) dx^\nu)+\mathfrak{G}_{\mu\nu}(r,x)dx^\mu dx^\nu  
\end{split}
\end{equation}
where  $\mathfrak{G}_{\mu\nu}$ is transverse, i.e., 
$u^\mu \mathfrak{G}_{\mu\nu}=0$. \footnote{All the Greek indices are raised and 
lowered using the boundary metric $g_{\mu\nu}$ defined by
\begin{equation}\label{eq:bndmet}
\begin{split}
g_{\mu\nu} = \lim_{r\rightarrow \infty} r^{-2} \left[ \mathfrak{G}_{\mu\nu}-  u_{(\mu} \mathcal{V}_{\nu)} \right]
\end{split}
\end{equation}
and $u_\mu$ is the unit time-like velocity field in the boundary, i.e., $g^{\mu\nu}u_\mu u_\nu = -1 $.}

For later purposes, we note that the inverse of this bulk metric takes the form
\begin{equation}\label{eq:invmetstd}
\begin{split}
u^\mu & \left[(\partial_\mu-\mathcal{V}_\mu\partial_r)\otimes\partial_r+\partial_r\otimes (\partial_\mu-\mathcal{V}_\mu\partial_r) \right]\\
&+(\mathfrak{G}^{-1})^{\mu\nu}(\partial_\mu-\mathcal{V}_\mu\partial_r)\otimes (\partial_\nu-\mathcal{V}_\nu\partial_r) \\
\end{split}
\end{equation}
where the symmetric matrix $(\mathfrak{G}^{-1})^{\mu\nu}$ is uniquely defined by the relations $u_{\mu}(\mathfrak{G}^{-1})^{\mu\nu}=0$ and $(\mathfrak{G}^{-1})^{\mu\lambda}\mathfrak{G}_{\lambda\nu}=\delta^{\mu}_\nu +u^\mu u_\nu\equiv P^\mu_\nu$.

Consider now a bulk-diffeomorphism of the form $r=e^{-\phi}\tilde{r}$ along with a scaling in the temperature of the form $b=e^{\phi}\tilde{b}$ where we assume that $\phi=\phi(x)$ is a function only of the boundary co-ordinates. The metric and the inverse metric components transform as
\begin{equation}\label{eq:bulkweyl}
\begin{split}
\mathcal{V}_{\mu} &= e^{-\phi}\left[\tilde{\mathcal{V}}_{\mu}+\tilde{r}\ \partial_\mu\phi \right],\quad\ u_\mu = e^{\phi}\tilde{u}_\mu, \quad  \mathfrak{G}_{\mu\nu}=\tilde{\mathfrak{G}}_{\mu\nu} \quad \text{and}\quad (\mathfrak{G}^{-1})^{\mu\nu}=(\tilde{\mathfrak{G}}^{-1})^{\mu\nu}\\
% dr+&\mathcal{V}_\nu dx^\nu = e^{-\phi} (d\tilde{r}+\tilde{\mathcal{V}}_\nu dx^\nu) \quad\text{and}\quad
% \partial_\mu-\mathcal{V}_\mu\partial_r= \tilde{\partial}_\mu-\tilde{\mathcal{V}}_\mu\tilde{\partial}_r \\
% \partial_r &= e^{\phi}\tilde{\partial}_r,  
\end{split}
\end{equation}
Recall however that, within our procedure, the quantities 
${\mathfrak{G}}_{\mu\nu}$ and $\mathcal{V}_{\mu}$ are each functions of 
$u^\mu$ and $b$. Now $u^\mu$ and $b$ each pick up a factor of $e^{\phi}$ 
under the same diffeomorphism (the transformation of $b$ 
is determined by examining the action of the diffeomorphism on 
\eqref{eqn:weylboostbrane}). We conclude that consistency demands that 
$\mathcal{V}_{\mu}$ and ${\mathfrak{G}}_{\mu\nu}$ are functions of $b$ 
and $u^\mu$ that respectively transform like a connection/remain
invariant under boundary Weyl transformation. It follows immediately 
that, for instance ${\mathfrak{G}}_{\mu\nu}$ is a linear sum of the 
Weyl invariant forms listed in section 2, with coefficients that are 
arbitrary functions of $br$. Similarly, $\mathcal{V}_\mu-r\mathcal{A}_\mu$ is 
a linear sum of Weyl-covariant vectors(both transverse and non-transverse) with
weight unity. 

Symmetry requirements do not constrain the form of these coefficients, 
which have to be determined via direct calculation. In the next section 
we simply present the results of such a calculation.

\section{The bulk metric and boundary stress tensor to second order} 
\subsection{The metric dual to hydrodynamics}

Using a Weyl-covariant form of the procedure outlined in \cite{Bhattacharyya:2008jc}, we find that the final metric can be written in the form 
\begin{equation}\label{metricsimp:eq}
\begin{split}
ds^2&=-2 u_\mu dx^\mu \left( dr + r\ A_\nu dx^\nu \right) + \left[ r^2 g_{\mu\nu} +u_{(\mu}\mathcal{S}_{\nu)\lambda}u^\lambda -\omega_{\mu}{}^{\lambda}\omega_{\lambda\nu}\right]dx^\mu dx^\nu\\
&+\frac{1}{(br)^d}(r^2-\frac{1}{2}\omega_{\alpha \beta}\omega^{\alpha \beta}) u_\mu u_\nu dx^\mu dx^\nu+2(br)^2 F(br)\left[\frac{1}{b}  \sigma_{\mu\nu} +  F(br)\sigma_{\mu}{}^{\lambda}\sigma_{\lambda \nu}\right]dx^\mu dx^\nu \\
&-2(br)^2 \left[K_1(br)\frac{\sigma_{\alpha \beta}\sigma^{\alpha \beta}}{d-1}P_{\mu\nu} + K_2(br)\frac{u_\mu u_\nu}{(br)^{d}}\frac{\sigma_{\alpha \beta}\sigma^{\alpha \beta}}{2(d-1)}-\frac{L(br)}{(br)^{d}} 
u_{(\mu}P_{\nu)}^{\lambda}\mathcal{D}_{\alpha}{\sigma^{\alpha}}_{\lambda}\right] dx^\mu dx^\nu\\
&-2(br)^2 H_1(br)\left[u^{\lambda}\mathcal{D}_{\lambda}\sigma_{\mu \nu}+\sigma_{\mu}{}^{\lambda}\sigma_{\lambda \nu} -\frac{\sigma_{\alpha \beta}\sigma^{\alpha \beta}}{d-1}P_{\mu \nu} + C_{\mu\alpha\nu\beta}u^\alpha u^\beta \right]dx^\mu dx^\nu \\
&+2(br)^{2} H_2(br)\left[u^{\lambda}\mathcal{D}_{\lambda}\sigma_{\mu \nu}+\omega_{\mu}{}^{\lambda}\sigma_{\lambda \nu}+\omega_\nu{}^\lambda \sigma_{\mu\lambda}\right]  dx^{\mu} dx^{\nu}\\
\end{split}
\end{equation}
We have checked using Mathematica that the above metric solves Einstein equations upto $d=10$.

The various functions appearing in the metric are defined by the integrals
\begin{equation*}\label{metricfns:eq}
\begin{split}
F(br)&\equiv \int_{br}^{\infty}\frac{y^{d-1}-1}{y(y^{d}-1)}dy \\
&\approx  \frac{1}{br} -\frac{1}{d(br)^d}+ \frac{1}{(d+1)(br)^{d+1}}+\frac{\#}{(br)^{2d}}+\ldots\\
\end{split}
\end{equation*}
\begin{equation*}
\begin{split}
H_1(br)&\equiv \int_{br}^{\infty}\frac{y^{d-2}-1}{y(y^{d}-1)}dy \\
&\approx \frac{1}{2(br)^2}-\frac{1}{d(br)^d}+ \frac{1}{(d+2)(br)^{d+2}}+\frac{\#}{(br)^{2d}}+\ldots\\  
\end{split}
\end{equation*}
\begin{equation*}
\begin{split}
H_2(br)&\equiv \int_{br}^{\infty}\frac{d\xi}{\xi(\xi^d-1)}
\int_{1}^{\xi}y^{d-3}dy \left[1+(d-1)y F(y) +2 y^{2} F'(y) \right]\\
&=\frac{1}{2} F(br)^2-\int_{br}^{\infty}\frac{d\xi}{\xi(\xi^d-1)}
\int_{1}^{\xi}\frac{y^{d-2}-1}{y(y^{d}-1)}dy\\
&\approx \frac{1}{2(br)^2}-\frac{1}{d(br)^d}\int_{1}^{\infty}\frac{y^{d-2}-1}{y(y^{d}-1)}dy\\
&\qquad -\frac{1}{d(br)^{d+1}}+\frac{3 d + 5}{2(d+1)(d+2)(br)^{d+2}}+\frac{\#}{(br)^{2d}}+\ldots \\
\end{split}
\end{equation*}
\begin{equation*}
\begin{split}
K_1(br) &\equiv \int_{br}^{\infty}\frac{d\xi}{\xi^2}\int_{\xi}^{\infty}dy\ y^2 F'(y)^2 \\
&\approx \frac{1}{2(br)^2}-\frac{2}{d(d+1)(br)^{d+1}}+\frac{2}{(d+1)(d+2)(br)^{d+2}}+\frac{\#}{(br)^{2d}}+\ldots\\
\end{split}
\end{equation*}
\begin{equation*}
\begin{split}
K_2(br) &\equiv \int_{br}^{\infty}\frac{d\xi}{\xi^2}\left[1-\xi(\xi-1)F'(\xi) -2(d-1)\xi^{d-1} \right.\\
&\left. \quad +\left(2(d-1)\xi^d-(d-2)\right)\int_{\xi}^{\infty}dy\ y^2 F'(y)^2 \right]\\
&\approx -\frac{(d-3)(d-1)}{2(d+1)(br)^2}+\frac{2(d-2)}{d(br)}+\frac{1}{d(2d-1)(br)^{d}}+\frac{\#}{(br)^{d+2}}+\ldots\\
\end{split}
\end{equation*}
\begin{equation*}
\begin{split}
L(br) &\equiv \int_{br}^\infty\xi^{d-1}d\xi\int_{\xi}^\infty dy\ \frac{y-1}{y^3(y^d
-1)} \\
&\approx -\frac{1}{d(d+2)(br)^2}+\frac{1}{(d+1)(br)}\\
&\qquad-\frac{1}{(d+1)(2d+1)(br)^{d+1}}-\frac{1}{2(d+1)(d+2)(br)^{d+2}} +\frac{\#}{(br)^{2d}}+\ldots \\
\end{split}
\end{equation*}
where we have also provided the asymptotic forms of the functions as their arguments tend to infinity.

Later in this paper, we will find it convenient to work with other equivalent forms of the above metric. Using
\begin{equation}\label{Sexp:eq}
\mathcal{S}_{\mu\lambda}u^\lambda=-\frac{1}{d-2}\mathcal{D}_\lambda \omega^{\lambda}{}_{\mu}+\frac{1}{d-2}\mathcal{D}_\lambda \sigma^{\lambda}{}_{\mu}
-\frac{\mathcal{R}}{2(d-1)(d-2)}u_{\mu}+\ldots
\end{equation}
we can write
\begin{equation}\label{GSexp:eq}
\begin{split}
ds^2&=-2 u_\mu dx^\mu \left( dr + r\ A_\nu dx^\nu \right) + r^2 g_{\mu\nu} dx^\mu dx^\nu\\
&-\left[\omega_{\mu}{}^{\lambda}\omega_{\lambda\nu}+\frac{1}{d-2}\mathcal{D}_\lambda \omega^{\lambda}{}_{(\mu}u_{\nu)}-\frac{1}{d-2}\mathcal{D}_\lambda \sigma^{\lambda}{}_{(\mu}u_{\nu)}
+\frac{\mathcal{R}}{(d-1)(d-2)}u_{\mu}u_{\nu}\right]dx^\mu dx^\nu\\ &+\frac{1}{(br)^d}(r^2-\frac{1}{2}\omega_{\alpha \beta}\omega^{\alpha \beta}) u_\mu u_\nu dx^\mu dx^\nu+2(br)^2 F(br)\left[\frac{1}{b}  \sigma_{\mu\nu} +  F(br)\sigma_{\mu}{}^{\lambda}\sigma_{\lambda \nu}\right]dx^\mu dx^\nu \\
&-2(br)^2 \left[K_1(br)\frac{\sigma_{\alpha \beta}\sigma^{\alpha \beta}}{d-1}P_{\mu\nu} + K_2(br)\frac{u_\mu u_\nu}{(br)^{d}}\frac{\sigma_{\alpha \beta}\sigma^{\alpha \beta}}{2(d-1)}-\frac{L(br)}{(br)^{d}} 
u_{(\mu}P_{\nu)}^{\lambda}\mathcal{D}_{\alpha}{\sigma^{\alpha}}_{\lambda}\right] dx^\mu dx^\nu\\
&-2(br)^2 H_1(br)\left[u^{\lambda}\mathcal{D}_{\lambda}\sigma_{\mu \nu}+\sigma_{\mu}{}^{\lambda}\sigma_{\lambda \nu} -\frac{\sigma_{\alpha \beta}\sigma^{\alpha \beta}}{d-1}P_{\mu \nu} + C_{\mu\alpha\nu\beta}u^\alpha u^\beta \right]dx^\mu dx^\nu \\
&+2(br)^{2} H_2(br)\left[u^{\lambda}\mathcal{D}_{\lambda}\sigma_{\mu \nu}+\omega_{\mu}{}^{\lambda}\sigma_{\lambda \nu}-\sigma_{\mu}{}^{\lambda}\omega_{\lambda \nu}\right]  dx^{\mu} dx^{\nu}\\
\end{split}
\end{equation}

or alternatively the metric can be written in the form \eqref{eq:metstd}

\begin{equation*}\label{eq:VGexp} 
\begin{split}
ds^2 &= -2 u_\mu dx^\mu(dr+\mathcal{V}_\nu dx^\nu)+\mathfrak{G}_{\mu\nu}dx^\mu dx^\nu \qquad \text{with}\\
\end{split}
\end{equation*}
\begin{equation*}
\begin{split}
\mathcal{V}_\mu &= r \mathcal{A}_\mu -\mathcal{S}_{\mu\lambda}u^\lambda-\frac{2 L(br)}{(br)^{d-2}} P_{\mu}^{\nu}\mathcal{D}_{\lambda}{\sigma^{\lambda}}_{\nu}\\ &-\frac{u_\mu}{2(br)^d}\left[
 r^2(1-(br)^d)-\frac{1}{2}\omega_{\alpha \beta}\omega^{\alpha \beta} 
-(br)^2 K_2(br)\frac{\sigma_{\alpha \beta}\sigma^{\alpha \beta}}{d-1}
\right]  +\ldots\\
&= r \mathcal{A}_\mu+\frac{1}{d-2}\left[\mathcal{D}_\lambda \omega^{\lambda}{}_{\mu}-\mathcal{D}_\lambda \sigma^{\lambda}{}_{\mu}
+\frac{\mathcal{R}}{2(d-1)}u_{\mu}\right]-\frac{2 L(br)}{(br)^{d-2}} P_{\mu}^{\nu}\mathcal{D}_{\lambda}{\sigma^{\lambda}}_{\nu}\\ 
& -\frac{u_\mu}{2(br)^d}\left[
 r^2(1-(br)^d)-\frac{1}{2}\omega_{\alpha \beta}\omega^{\alpha \beta} 
-(br)^2 K_2(br)\frac{\sigma_{\alpha \beta}\sigma^{\alpha \beta}}{d-1}
\right]  +\ldots\\
\end{split}
\end{equation*}
\begin{equation}
\begin{split}
\mathfrak{G}_{\mu\nu}&= r^2 P_{\mu\nu}-\omega_{\mu}{}^{\lambda}\omega_{\lambda\nu}\\
&+2(br)^2 F(br)\left[\frac{1}{b}  \sigma_{\mu\nu} +  F(br)\sigma_{\mu}{}^{\lambda}\sigma_{\lambda \nu}\right]-2(br)^2 K_1(br)\frac{\sigma_{\alpha \beta}\sigma^{\alpha \beta}}{d-1}P_{\mu\nu}\\
&-2(br)^2 H_1(br)\left[u^{\lambda}\mathcal{D}_{\lambda}\sigma_{\mu \nu}+\sigma_{\mu}{}^{\lambda}\sigma_{\lambda \nu} -\frac{\sigma_{\alpha \beta}\sigma^{\alpha \beta}}{d-1}P_{\mu \nu} + C_{\mu\alpha\nu\beta}u^\alpha u^\beta \right]\\
&+2(br)^{2} H_2(br)\left[u^{\lambda}\mathcal{D}_{\lambda}\sigma_{\mu \nu}+\omega_{\mu}{}^{\lambda}\sigma_{\lambda \nu}+\omega_\nu{}^\lambda \sigma_{\mu\lambda}\right] +\ldots\\
\end{split}
\end{equation}

Using \eqref{eq:invmetstd}, the inverse metric can be calculated . The tensor $(\mathfrak{G}^{-1})^{\mu\nu}$ occurring in the inverse metric can be calculated as
\begin{equation}\label{eq:Ginv}
\begin{split}
(\mathfrak{G}^{-1})^{\mu\nu}&= \frac{1}{r^2} P^{\mu\nu}+\frac{1}{r^4}\omega^{\mu\lambda}\omega_{\lambda}{}^{\nu}\\
&-\frac{2 b^2}{r^2} F(br)  \left[\frac{1}{b}  \sigma^{\mu\nu} -  F(br)\sigma^{\mu}{}_{\lambda}\sigma^{\lambda \nu}\right] +\frac{2 b^2}{r^2} K_1(br)\frac{\sigma_{\alpha \beta}\sigma^{\alpha \beta}}{d-1}P^{\mu\nu}\\
&+\frac{2 b^2}{r^2}H_1(br)\left[u^{\lambda}\mathcal{D}_{\lambda}\sigma^{\mu \nu}+\sigma^{\mu}{}_{\lambda}\sigma^{\lambda \nu} -\frac{\sigma_{\alpha \beta}\sigma^{\alpha \beta}}{d-1}P^{\mu \nu} + C^{\mu\alpha\nu\beta}u_\alpha u_\beta \right]\\
&-\frac{2 b^2}{r^2} H_2(br)\left[u^{\lambda}\mathcal{D}_{\lambda}\sigma^{\mu \nu}+\omega^{\mu}{}_{\lambda}\sigma^{\lambda \nu}+\omega^\nu{}_\lambda \sigma^{\mu\lambda}\right] +\ldots
\end{split}
\end{equation}
 
We have checked that results of this subsection agree with the hydrodynamic 
metric duals for $d=4$ derived by the authors of \cite{Bhattacharyya:2008jc} and the $d=3$ metric derived in \cite{VanRaamsdonk:2008fp} (in order to match our 
results with older work that was performed in different gauges we implemented 
the necessary gauge transformations). In the next subsection, we proceed 
to derive the stress tensor dual to this metric and compare it against the results available in the literature.

\subsection{Energy momentum tensor of fluids with metric duals}

The dual stress tensor corresponding to the metric in the previous subsection is given by 
\begin{equation}\label{enmom:eq}
\begin{split}
T_{\mu\nu} &= p\left(g_{\mu\nu}+d u_\mu u_\nu \right)-2\eta \sigma_{\mu\nu}\\
&-2\eta \tau_\omega \left[u^{\lambda}\mathcal{D}_{\lambda}\sigma_{\mu \nu}+\omega_{\mu}{}^{\lambda}\sigma_{\lambda \nu}+\omega_\nu{}^\lambda \sigma_{\mu\lambda} \right]\\
&+2\eta b\left[u^{\lambda}\mathcal{D}_{\lambda}\sigma_{\mu \nu}+\sigma_{\mu}{}^{\lambda}\sigma_{\lambda \nu} -\frac{\sigma_{\alpha \beta}\sigma^{\alpha \beta}}{d-1}P_{\mu \nu}+ C_{\mu\alpha\nu\beta}u^\alpha u^\beta \right]\\
\end{split}
\end{equation}
with
\begin{equation}\label{enparam:eq}
\begin{split}
b=\frac{d}{4\pi T}\qquad;&\qquad p=\frac{1}{16\pi G_{\text{AdS}}b^d}\\
\eta = \frac{s}{4\pi}=\frac{1}{16\pi G_{\text{AdS}}b^{d-1}}\quad
\text{and}& \qquad \tau_{\omega} =  b \int_{1}^{\infty}\frac{y^{d-2}-1}{y(y^{d}-1)}dy 
\end{split}
\end{equation}
This result is a generalization to the fluid dynamical stress tensor 
on an arbitrary curved manifold in general dimension $d$ reported in
\cite{Baier:2007ix, Bhattacharyya:2008jc,Loganayagam:2008is,
VanRaamsdonk:2008fp, Natsuume:2008iy, Bhattacharyya:2008ji} for special 
values of $d$ and most recently by \cite{Haack:2008cp} for flat space 
in arbitrary dimensions. The values of $\tau_\omega$ for some of the lower 
dimensions is shown\footnote{More generally, the integral appearing in 
the expression for $\tau_\omega$ can be evaluated in terms of the derivative
of the Gamma function or more directly in terms of `the harmonic number
function' with the fractional argument(as was noted in \cite{Natsuume:2008gy})
\[\tau_\omega = -\frac{b}{d}\left[\gamma_E+\frac{d}{dz}\text{Log}\ \Gamma(z)\right]_{z=2/d} = -\frac{b}{d} \text{Harmonic}[2/d-1] \] For large $d$, $\tau_\omega$ has an expansion of the form $\tau_\omega/b=1/2-\pi^2/(3d^2)+\ldots$.}
 in the table~\ref{tauomega:tab}.

Comparing this expression with \eqref{Tgeneral:eq}, we get
\begin{equation}\label{Tgeneralcoeff:eq}
\begin{split}
\xi_\sigma = \xi_C = 2\eta(\tau_1+\tau_2) = 2\eta b \quad,\quad \tau_2=\tau_{\omega}\quad &\text{and}\quad \xi_\omega =0 \\ 
\end{split}
\end{equation}
Note that these relations between $\xi_\sigma ,\xi_C,\tau_1$ and $\tau_2$ quoted above are universal in the sense that they hold true for uncharged fluids in arbitrary dimensions with the gravity duals. It would be interesting to check whether these relations between the transport coefficients continue to hold against various possible generalizations including the generalization to fluids with one or more global conserved charge.\footnote{In this context, we would like to note that in the presence of a charge, there are more than one natural convention for the definition of the velocity - velocity can be defined as the unit time-like eigenvector of the energy-momentum tensor (as we have done in the chargeless case) or can be defined alternatively to be the unit time-like vector along the charge current. The former is called the Landau frame velocity and the latter is termed the velocity in the Eckart frame. The transport coefficients defined above can depend crucially on which of these definitions are used.

While this work was nearing completion, the authors of \cite{Erdmenger:2008rm} and \cite{Banerjee:2008th} reported independently the transport coefficients for a particular class of charged black brane configurations with flat boundaries. Interestingly, their coefficients continue to obey $\xi_\sigma=2\eta(\tau_1 + \tau_2)$ (or equivalently $4\lambda_1+\lambda_2=2\eta \tau_\Pi$) in the Landau frame. As far as we know, the charge dependence of $\xi_C$ is not known yet. Authors of \cite{Erdmenger:2008rm} and \cite{Banerjee:2008th} report $\xi_\omega \neq 0$ for the charged case in the Landau frame.} 

Now, we proceed to compare our results against the results already available in the literature. Until now, we have found it convenient to closely follow the parametrisation of the stress tensor in \cite{Loganayagam:2008is}. An alternative parametrisation of the energy-momentum tensor was presented in the section 3.1 of \cite{Baier:2007ix} - the parameters $\tau_{_{\Pi}},\lambda_{1,2,3}$ and $\kappa$ defined there can be related to our parameters via the relations
\begin{equation}\label{baierrelation:eq}
\begin{split}
\tau_1 = \tau_{_{\Pi}}\quad,\quad \tau_2 = -\frac{\lambda_2}{2\eta} \quad,\quad
\xi_\sigma = 4\lambda_1 \quad,\quad \xi_C = \kappa(d-2) \quad\text{and}\quad \xi_\omega =\lambda_3 \\ 
\end{split}
\end{equation}
which in turn gives the value of the transport coefficients as 
\begin{equation}\label{baiercoeff:eq}
\begin{split}
\tau_{\Pi}=b-\tau_\omega \quad,\quad \lambda_1=\frac{\eta b}{2} \quad,\quad \lambda_2=-2\eta \tau_\omega\quad,\quad \lambda_3=0 \quad\text{and}\quad \kappa=\frac{2\eta b}{d-2}
\end{split}
\end{equation}
which agrees with all the previous results in the literature \cite{Natsuume:2008iy,VanRaamsdonk:2008fp,Haack:2008cp}.

\begin{table}\label{tauomega:tab}
  \centering
  \begin{tabular}{||r|l|l||}
    \hline
    \multicolumn{3}{||c||}{Value of $\tau_{\omega}/b$ for various dimensions }\\
    \hline
    $d$ & Value of $\tau_{\omega}/b=\int_{1}^{\infty}\frac{y^{d-2}-1}{y(y^{d}-1)}dy $ & $\tau_{\omega}/b$ (Numerical)\\
    \hline
    3 & $\frac{1}{2}\left(\text{Log}\ 3-\frac{\pi}{3\sqrt{3}}\right)$ & 0.247006\ldots \\
    4 & $\frac{1}{2}\text{Log}\ 2$                                    & 0.346574\ldots  \\
    5 & $\frac{1}{4}\left(\text{Log}\ 5 + \frac{2\pi}{5} \sqrt{1-\frac{2}{\sqrt{5}}} -\frac{2}{\sqrt{5}} \text{ArcCoth}\ \sqrt{5} \right)$ & 0.396834\ldots\\
    6 & $\frac{1}{4}\left(\text{Log}\ 3+\frac{\pi}{3\sqrt{3}}\right)$ & 0.425803\ldots \\
%     7 & & \\
%     8 & & \\
%     9 & & \\
%    10 & & \\
    \hline
\end{tabular}
\end{table}

%_____________________________________________
\section{Causal structure and the local entropy current}
\label{entropy}
%_____________________________________________

\subsection{The event horizon of our solutions}

Although our assumptions can almost certainly be greatly relaxed, for 
the purposes of this section we specialize to boundary metrics that 
settle down, at late times to either the flat metric on $R^{d-1,1}$ or the 
flat metric on $S^{d-1} \times $ time and to fluid flows that settle down
at late times to uniform brane configurations on $R^{d-1,1}$ or stationary 
rotating black holes (studied in greater detail ahead) on  $S^{d-1} \times $
time. See \cite{Bhattacharyya:2008xc} for a discussion on how the dissipative
nature of fluid dynamics makes the last assumption less restrictive than 
it naively seems. 

Now the event horizon of our spacetimes is simply the unique null hypersurface
that tends, at late times, to the known event horizons of the late time limit of our solutions. In this subsection we will explain how this clear characterization may be translated into an explicit and local mathematical formula for the event horizon within the derivative expansion.

Recall that our bulk metric is written in the gauge $g_{rr}=0$, 
$g_{r\mu}=-u_\mu$, and consequently takes the form 
\begin{equation}\label{eq:metstd1}
\begin{split}
ds^2 &= -2 u_\mu dx^\mu(dr+\mathcal{V}_\nu dx^\nu)+\mathfrak{G}_{\mu\nu}dx^\mu dx^\nu  
\end{split}
\end{equation}
where we remind the reader that $\mathfrak{G}_{\mu\nu}$ is transverse and all the Greek indices are raised using the boundary metric $g_{\mu\nu}$. As we 
have explained before $\mathcal{V}_\mu$ transforms like a connection and 
$\mathfrak{G}_{\mu\nu}$ is invariant under boundary Weyl transformations. 

Let us suppose that the event horizon is given by the equation 
$\mathcal{S}\equiv r-\rH(x)=0$. The normal vector $\xi_A$ to this hypersurface 
is simply the one-form $dS= \xi_A dy^A = dr -\partial_\mu r_H dx^\mu$. This one-form - and 
its dual normal vector - can be written in a manifestly Weyl covariant 
(if slightly complicated) form as follows
\begin{equation}\label{eq:normstd}
\begin{split}
\xi_A dy^A &= dS= (dr+\mathcal{V}_\lambda dx^\lambda)-\kappa_\mu dx^\mu\\
\xi^A\partial_A &={G}^{AB}\partial_{A}\mathcal{S}\partial_{B}= n^\mu(\partial_\mu-\mathcal{V}_\mu\partial_r)-u^\mu\kappa_\mu\partial_r\\
&= n^\mu\left[\partial_\mu+\partial_\mu\rH \partial_r\right] =n^\mu\left[\partial_\mu\right]_{r=\rH}
\end{split}
\end{equation}
where we have introduced two new Weyl-covariant vectors $\kappa^\mu=e^{-\phi}\tilde{\kappa}^\mu$ and $n^\mu=e^{-\phi}\tilde{n}^\mu$ defined via
\begin{equation}
\begin{split}
\kappa_\mu &\equiv \partial_\mu \rH+\mathcal{V}_{\mu H} \qquad \text{and}\\
n^\mu &\equiv u^\mu-(\mathfrak{G}_H^{-1})^{\mu\nu}\kappa_\nu \\
\end{split}
\end{equation}
We use the subscript $H$ to denote that the functions are to be evaluated at the event-horizon.

If we adopt the boundary co-ordinates $x^\mu$ as the co-ordinates on the event horizon, the induced metric on the horizon can be written as
\begin{equation}\label{eq:HorInd}
\begin{split}
ds^2_H = \left[G_{AB}(y) dy^A dy^B\right]_{r=\rH(x)} \equiv \mathcal{H}_{\mu\nu}(x) dx^\mu dx^\nu
\end{split}
\end{equation}
with
\begin{equation}\label{eq:Hdef}
\begin{split}
\mathcal{H}_{\mu\nu}= \mathfrak{G}_{\mu\nu}-u_{(\mu}\kappa_{\nu)}
\end{split}
\end{equation}
and the null-condition on the horizon, $[G_{AB}]_H \xi^A \xi^B=\mathcal{H}_{\mu\nu}n^\mu n^\nu=0$ translates to
\begin{equation}\label{eq:nullcondn}
(\mathfrak{G}^{-1})^{\mu\nu}\kappa_\mu\kappa_\nu = 2 u^\mu\kappa_\mu
\end{equation}

We now follow  \cite{Bhattacharyya:2008ji} to compute the event horizon of 
our solutions in the derivative expansion. 
We start from a Weyl-covariant derivative expansion for $\rH$ given by
\begin{equation}\label{eq:rHexp}
\begin{split}
\rH &= \frac{1}{b}+b \left(h_1 \sigma_{\alpha\beta} \sigma^{\alpha\beta} + h_2 \omega_{\alpha\beta} \omega^{\alpha\beta} +h_3 \mathcal{R} \right) +\ldots\\
&= \rH^{(0)} + \rH^{(2)}+\ldots
\end{split}
\end{equation}
Note that, since there is no first order Weyl-covariant scalar,\footnote{See \cite{Bhattacharyya:2008ji} for a classification of the possible Weyl-covariant tensors.} there are no corrections to $\rH$ at the first order in the derivative expansion.

We first compute $\kappa_\mu$
\begin{equation}\label{kappasol:eq}
\begin{split}
\kappa_\mu &= \mathcal{D}_\mu b^{-1} - \mathcal{S}_{\mu\lambda} u^\lambda -2 L_H P_{\mu}^{\nu}\mathcal{D}_{\lambda}{\sigma^{\lambda}}_{\nu}   \\
& + u_\mu \left[
 \frac{1}{4}\omega_{\alpha \beta}\omega^{\alpha \beta} 
+ \frac{K_{2H}}{2(d-1)}\sigma_{\alpha \beta}\sigma^{\alpha \beta}+\frac{d}{2b}\rH^{(2)}
\right] +\ldots\\
% n^\mu &= u^\mu - b^2 P^{\mu\nu}\left[\mathcal{D}_\nu b^{-1} -\mathcal{S}_{\nu\lambda} u^\lambda -2 L_H \mathcal{D}_{\lambda}{\sigma^{\lambda}}_{\nu}\right]\\
% \sqrt{\text{det}_{d-1}\mathfrak{G}_H} &= \frac{1}{b^{d-1}}\left[1+(d-1)b\rH^{(2)}+
% \frac{b^2}{2}\omega_{\alpha \beta}\omega^{\alpha \beta}-b^2 K_{1H} \sigma_{\alpha \beta}\sigma^{\alpha \beta}\right]\\
% b^{d-1} n^\mu \sqrt{\text{det}_{d-1}\mathfrak{G}_H} &=\left[1+(d-1)b\rH^{(2)}+
% \frac{b^2}{2}\omega_{\alpha \beta}\omega^{\alpha \beta}-b^2 K_{1H} \sigma_{\alpha \beta}\sigma^{\alpha \beta} \right]u^\mu\\
% &- b^2 P^{\mu\nu}\left[\mathcal{D}_\nu b^{-1} - \mathcal{S}_{\nu\lambda} u^\lambda -2 L_H \mathcal{D}_{\lambda}{\sigma^{\lambda}}_{\nu}\right]\\
\end{split}
\end{equation}

Substituting the above into \eqref{eq:nullcondn}, we get
\begin{equation}\label{rHsol:eq}
\begin{split}
\rH^{(2)} &= \frac{2b}{d}\left[u^\mu(\mathcal{D}_\mu b^{-1} 
 - \mathcal{S}_{\mu\nu}u^\nu) 
- \frac{1}{4}\omega_{\alpha \beta}\omega^{\alpha \beta} 
- \frac{K_{2H}}{2(d-1)}\sigma_{\alpha \beta}\sigma^{\alpha \beta}\right]\\
\end{split}
\end{equation}
% \begin{equation}
% \begin{split}
% b^{d-1} n_\mu \sqrt{\text{det}_{d-1}\mathfrak{G}_H} &=u_\mu- b^2 P_{\mu}^{\nu}\left[\mathcal{D}_\nu b^{-1} - \mathcal{S}_{\nu\lambda} u^\lambda -2 L_H \mathcal{D}_{\lambda}{\sigma^{\lambda}}_{\nu}\right]+\frac{b^2 u_\mu}{2d}\omega_{\alpha \beta}\omega^{\alpha \beta}\\
% &+b^2 u_\mu\left[\frac{2(d-1)}{d}u^\alpha(\mathcal{D}_\alpha b^{-1}  - \mathcal{S}_{\alpha\beta}u^\beta)
% - (K_{1H}+\frac{K_{2H}}{d})\sigma_{\alpha \beta}\sigma^{\alpha \beta}\right]\\
% &=u_\mu- b^2 \left[\mathcal{D}_{\mu} b^{-1} - \mathcal{S}_{\mu\lambda} u^\lambda -2 L_H \mathcal{D}_{\lambda}{\sigma^{\lambda}}_{\mu}\right]+\frac{b^2 u_\mu}{2d}\omega_{\alpha \beta}\omega^{\alpha \beta}\\
% &+b^2 u_\mu\left[\frac{d-2}{d}u^\alpha(\mathcal{D}_\alpha b^{-1} 
%  - \mathcal{S}_{\alpha\beta}u^\beta)-(K_{1H}+\frac{K_{2H}}{d}+2 L_H)\sigma_{\alpha \beta}\sigma^{\alpha \beta}\right]
% \end{split}
% \end{equation}

To bring this to the form \eqref{eq:rHexp}, we use \eqref{DTcons:eq} and \eqref{Sexp:eq} to write 
\begin{equation*}
\begin{split}
\mathcal{D}_\mu b^{-1}-\mathcal{S}_{\mu\nu}u^\nu &= \left(\frac{2}{d}-\frac{1}{d-2}\right) \mathcal{D}_{\lambda} \sigma^{\lambda}{}_{\mu}+\frac{\mathcal{D}_{\lambda} \omega^{\lambda}{}_{\mu}}{d-2}\\
&- \frac{2}{d-1} \sigma_{\alpha\beta} \sigma^{\alpha\beta}u_\mu+\frac{\mathcal{R} u_\mu}{2(d-1)(d-2)}+\ldots\\
\end{split}
\end{equation*}
and
\begin{equation}\label{LHeq}
\begin{split}
% K_{1H} d+ K_{2H}&= \int_{1}^{\infty}\frac{d\xi}{\xi^2}\left[1-\xi(\xi-1)F'(\xi) -2(d-1)\xi^{d-1} \right.\\
% &\left. \quad +2\left((d-1)\xi^d+1\right)\int_{\xi}^{\infty}dy\ y^2 F'(y)^2 \right]\\
L_H &= \int_{1}^\infty\xi^{d-1}d\xi\int_{\xi}^\infty dy\ \frac{y-1}{y^3(y^d-1)}=\frac{1}{2d} 
\end{split}
\end{equation}
which gives us the position of the event horizon as
\begin{equation}\label{eq:rHfinal}
\begin{split}
\rH &= \frac{1}{b}+b \left(h_1 \sigma_{\alpha\beta} \sigma^{\alpha\beta} + h_2 \omega_{\alpha\beta} \omega^{\alpha\beta} +h_3 \mathcal{R} \right) +\ldots\\
\end{split}
\end{equation}
where
\begin{equation}
\begin{split}
h_1&=\frac{2(d^2+d-4)}{d^2(d-1)(d-2)} - \frac{K_{2H}}{d(d-1)}\\
h_2&=-\frac{d+2}{2d(d-2)} \ \qquad \text{and}\quad h_3=-\frac{1}{d(d-1)(d-2)}\\
&\text{with} \qquad K_{2H} = \int_{1}^{\infty}\frac{d\xi}{\xi^2}\left[1-\xi(\xi-1)F'(\xi) -2(d-1)\xi^{d-1} \right.\\
 &\left. \quad +2\left((d-1)\xi^d-(d-2)\right)\int_{\xi}^{\infty}dy\ y^2 F'(y)^2 \right]
\end{split}
\end{equation}

\subsection{Entropy current as the pullback of Area form}

Once the event-horizon is obtained, one can compute an area form on the horizon which when pulled-back to boundary along the ingoing null geodesics gives the entropy current.This general prescription by \cite{Bhattacharyya:2008xc} translates into the following expression for the boundary entropy current\footnote{A more detailed justification of this formula can be found in \cite{Bhattacharyya:2008xc}}
\begin{equation}\label{Jseq}
\begin{split}
J^\mu_S &= \frac{\sqrt{\text{det}^{(n)}_{d-1}\mathcal{H}}}{4 G_{AdS}}n^\mu \\
&= \frac{\sqrt{\text{det}^{(n)}_{d-1}\mathcal{H}}}{4 G_{AdS}}\left[u^\mu-(\mathfrak{G}_H^{-1})^{\mu\nu}\kappa_\nu\right]
\end{split}
\end{equation}
where we will define $\text{det}^{(n)}_{d-1}\mathcal{H}$ in the following.

To define $\text{det}^{(n)}_{d-1}\mathcal{H}$ we will split the boundary co-ordinates $x^\mu$ to $(v,x^i)$ and we continue to use the same co-ordinates also on the event horizon. Under this split, the components of the $n^\mu$ also spilt into $(n^v,n^i)$. We will denote the $d-1$ dimensional induced metric on the constant $v$ submanifolds of the event horizon by $\mathfrak{h}_{ij}$.Then, we define
\begin{equation}
\begin{split}
\sqrt{\text{det}^{(n)}_{d-1}\mathcal{H}} = \frac{\sqrt{\text{det}_{d-1}\mathfrak{h}}}{n^v \sqrt{-\text{det} g}}
\end{split}
\end{equation} 
where $g_{\mu\nu}$ is the boundary metric and the expression on the right hand side has been assumed to be pulled back from the horizon to the boundary via the ingoing null-geodesics. Though we have used a particular split to define $\text{det}^{(n)}_{d-1}\mathcal{H}$, it can be shown that the answer that we get in the end is independent of the split(See section 3.3 of \cite{Bhattacharyya:2008xc}). Hence,the expression in \eqref{Jseq} constitutes a specific proposal for what the entropy current of the boundary fluid should be. This construction has the advantage that the second law in the boundary theory is automatically guaranteed by the area increase theorem in the bulk.\footnote{The Area increase theorem 
states that under appropriate assumptions the area of a blackhole can never 
decrease. This statement was proved by Hawking for the case of asymptotically 
flat spacetimes and is by now standard text book material 
(see e.g. \cite{Hawking:1973uf,1984ucp..book.....W}). This theorem has 
since been extended to black holes in more general spacetimes 
(see e.g. \cite{1982GReGr..14..793K,1984GReGr..16..121K,1984GReGr..16..365K}), 
including asymptotically $AdS$ spaces (see \cite{Chrusciel:2000cu} and 
references therein for a clear statement to this effect).}

% --------------------------------------
% where we remind the reader that $\text{det}_{d-1}\mathfrak{G}$ is defined via
% \begin{equation}\label{eq:detGdefn2}
% \begin{split}
% u^\lambda\epsilon_{\lambda\mu\nu\ldots}\ \mathfrak{G}^{\mu}_{\alpha}\  \mathfrak{G}^\nu_\beta\ \ldots = \left[\text{det}_{d-1}\mathfrak{G}\right]\  u^\lambda\epsilon_{\lambda\alpha\beta\ldots}
% \end{split}
% \end{equation}
% 
% 
% \begin{equation}\label{eq:detG}
% \begin{split}
% \sqrt{\text{det}_{d-1}\mathfrak{G}} &= r^{d-1}\left[1+
% \frac{\omega_{\alpha \beta}\omega^{\alpha \beta}-2(br)^2 K_1(br) \sigma_{\alpha \beta}\sigma^{\alpha \beta}}{2 r^2}+\ldots \right] \\
% \end{split}
% \end{equation}
% -------------------------------------------

By following the procedure just outlined, the dual entropy current of the conformal fluid can be calculated. We get 
\begin{equation}\label{Jsfinal:eq}
\begin{split}
4G_{AdS}b^{d-1} J^\mu_S &=u^\mu +b^2 u^\mu\left[A_1\ \sigma_{\alpha\beta}\sigma^{\alpha\beta}+A_2\  \omega_{\alpha\beta}\omega^{\alpha\beta}+ A_3\  \mathcal{R} \right]\\
&\qquad +b^2\left[B_1\ \mathcal{D}_\lambda\sigma^{\mu\lambda} + B_2\ \mathcal{D}_\lambda\omega^{\mu\lambda} \right]+\ldots \\
\end{split}
\end{equation}
with
\begin{equation}\label{eq:JSparam}
\begin{split}
A_1 &= \frac{2}{d^2}(d+2)-\frac{K_{1H} d+K_{2H}}{d}\ , \quad\ A_2 = -\frac{1}{2d} , \quad\ B_2 = \frac{1}{d-2}\\
B_1&= -2 A_3 =\frac{2}{d(d-2)}
\end{split}
\end{equation}
where $K_{1H} d+ K_{2H}$ is given by the integral
\begin{equation}\label{KHeq}
\begin{split}
K_{1H} d+ K_{2H}&= \int_{1}^{\infty}\frac{d\xi}{\xi^2}\left[1-\xi(\xi-1)F'(\xi) -2(d-1)\xi^{d-1} \right.\\
&\left. \quad +2\left((d-1)\xi^d+1\right)\int_{\xi}^{\infty}dy\ y^2 F'(y)^2 \right]\\
\end{split}
\end{equation}

% \begin{equation}
% \begin{split}
% J^\mu_S &= \frac{u^\mu}{4G_{AdS}b^{d-1}}+\frac{1}{4G_{AdS}b^{d-3}} u^\mu\left[A_1  \sigma_{\alpha\beta}\sigma^{\alpha\beta}-\frac{1}{2d}  \omega_{\alpha\beta}\omega^{\alpha\beta}-\frac{1}{d(d-2)} \mathcal{R} \right]\\
% &+\frac{1}{4G_{AdS}b^{d-3}}\left[\frac{2}{d(d-2)} \mathcal{D}_\lambda\sigma^{\mu\lambda} +\frac{1}{d-2} \mathcal{D}_\lambda\omega^{\mu\lambda} \right]+\ldots \\
% \text{with}& \\
% A_1 &= \frac{2}{d^2}(d+2)-(K_{1H}+\frac{K_{2H}}{d}) \\
% \end{split}
% \end{equation}

%____________________________________________
\subsection{Second law and the Rate of entropy production}
%_____________________________________________

In the absence of a clear field theoretic microscopic definition, 
it may be pragmatic to regard the entropy current of fluid dynamics 
as any local functional 
of the fluid dynamical variables whose divergence is non negative on every 
solution to the equations of motion of fluid dynamics, and which integrates
to the thermodynamic notion of entropy in equilibrium. According to this  
characterization the entropy current is {\it any} local Boltzmann $H$ 
function, whose monotonic increase characterizes the dissipative 
irreversibility of fluid flows.

In the previous subsection we have used the dual bulk description to 
give a `natural' bulk definition of the entropy current that satisfies 
all these properties. However, at least at the two derivative level, the 
construction of the previous subsection is not the unique 
construction that satisfies the requirements spelt out in the paragraph above. 

In this subsection we will take a purely algebraic approach to determine the 
most general Weyl covariant two derivative entropy current that has a non 
negative divergence, given the equations of motion derived above. The 
entropy current of the previous section will turn out to be one of 
a 4-parameter class of solutions to this constraint.  

The most general entropy current consistent with Weyl covariance\footnote{ We assume that there are no pseudo-vector contributions to the entropy current which can possibly appear only in the case of $d=4$. See \cite{Bhattacharyya:2008ji} for an analysis in $d=4$ including pseudovectors.} can be written as  
\begin{equation}\label{Jsgeneral:eq}
\begin{split}
4G_{\text{AdS}}b^{d-1} J^\mu_S &= u^\mu + b^2 u^\mu\left[A_1 \sigma_{\alpha\beta}\sigma^{\alpha\beta}+A_2  \omega_{\alpha\beta}\omega^{\alpha\beta}+ A_3  \mathcal{R} \right]\\
&+b^2 \left[B_1 \mathcal{D}_\lambda\sigma^{\mu\lambda} +B_2 \mathcal{D}_\lambda\omega^{\mu\lambda} \right]+\ldots \\
\end{split}
\end{equation}
Since we want to constrain the entropy current upto second order we will need to calculate the divergence of this current. In order to perform the calculation
in a Weyl covariant fashion we note that the ordinary divergence 
$\nabla_\mu J^\mu_S$ can be replaced by the Weyl-covariant divergence $ \mathcal{D}_\mu J^\mu_S $ with
\begin{equation}\label{DJdef:eq}
\begin{split}
\mathcal{D}_\mu J^\mu_S \equiv \nabla_\mu J^\mu_S +(w-d)\mathcal{A}_\mu J^\mu_S
\end{split}
\end{equation}
as the conformal weight of any entropy current must be $d$. 

Let us now take the divergence of \eqref{Jsgeneral:eq}. We find 
\begin{equation}\label{DdotJs:eq}
\begin{split}
4G_{\text{AdS}}b^{d-1} \mathcal{D}_\mu J^\mu_S &= (d-1)\ b\ u^\mu\mathcal{D}_\mu b^{-1} +  b^2 u^\mu \mathcal{D}_\mu\left[A_1 \sigma_{\alpha\beta}\sigma^{\alpha\beta}+A_2  \omega_{\alpha\beta}\omega^{\alpha\beta}+ A_3  \mathcal{R} \right]\\
&\qquad+ b^2 \mathcal{D}_\mu\left[B_1 \mathcal{D}_\lambda\sigma^{\mu\lambda} +B_2 \mathcal{D}_\lambda\omega^{\mu\lambda} \right]+\ldots \\
\end{split}
\end{equation}
which can in turn be evaluated using the following identities:\footnote{The first of these identities is just the re-statement of energy conservation $u_\mu\mathcal{D}_\nu T^{\mu\nu}=0$. The rest of them can be obtained by exploiting the properties of various Weyl-covariant quantities which are detailed in  \cite{Loganayagam:2008is}. Note however that the curvature tensors used here are the negative of those appearing in \cite{Loganayagam:2008is}.} 
\begin{equation}
\label{iden}
\begin{split}
(d-1)\ b\ u^\mu\mathcal{D}_\mu b^{-1}&= -\frac{\sigma_{\mu\nu}T^{\mu\nu}}{pd} \\
u^\mu \mathcal{D}_\mu\left[ \sigma_{\alpha\beta}\sigma^{\alpha\beta}\right] &=  2 \sigma_{\mu\nu} u^\lambda \mathcal{D}_\lambda\sigma^{\mu\nu} \\
u^\mu \mathcal{D}_\mu\left[\omega_{\alpha\beta}\omega^{\alpha\beta} \right] &= 4 \sigma_{\mu\nu} \omega^{\mu\lambda}\omega_\lambda{}^\nu + \omega_{\mu\nu} \mathcal{F}^{\mu\nu} \\
u^\mu \mathcal{D}_\mu\mathcal{R}  &= -2 \sigma^{\mu\nu} \mathcal{R}_{\mu\nu} + \omega_{\mu\nu} \mathcal{F}^{\mu\nu} +2 \mathcal{D}_\mu \mathcal{D}_\nu \sigma^{\mu\nu} - 2(d-2) \mathcal{D}_\mu \left[u_\nu\mathcal{F}^{\mu\nu}\right] \\
&= -2 (d-2)\sigma^{\mu\nu} \left[\sigma_{\mu}{}^{\lambda}\sigma_{\lambda \nu} +\omega_{\mu\lambda}\omega^\lambda{}_\nu + u^{\lambda}\mathcal{D}_{\lambda}\sigma_{\mu \nu}+C_{\mu\alpha\nu\beta}u^\alpha u^\beta \right] \\
&\qquad + \omega_{\mu\nu} \mathcal{F}^{\mu\nu} +2 \mathcal{D}_\mu \mathcal{D}_\nu \sigma^{\mu\nu} - 2(d-2) \mathcal{D}_\mu \left[u_\nu\mathcal{F}^{\mu\nu}\right] \\
\mathcal{D}_\mu \mathcal{D}_\nu\omega^{\mu\nu} &= -\frac{(d-3)}{2}\ \omega_{\mu\nu} \mathcal{F}^{\mu\nu}
\end{split}
\end{equation}

Substituting for the energy-momentum tensor from \eqref{enmom:eq} and keeping only those terms which contain no more than three derivatives, we get\footnote{Note, in particular, that $\mathcal{F}_{\mu\nu}$ is zero on-shell upto second order in the derivative expansion(See \eqref{integb:eq}).}
\begin{equation}
\begin{split}
4G_{\text{AdS}}b^{d-1} \mathcal{D}_\mu J^\mu_S &= \frac{2b}{d} \sigma^{\mu\nu}\left[ \sigma_{\mu\nu} -bd(d-2)\left(A_3 -\frac{2 A_2}{d-2} \right) \omega_{\mu\lambda}\omega^\lambda{}_\nu\right.\\
&-bd(d-2)\left(A_3 + \frac{1}{d(d-2)}\right)\left(\sigma_{\mu}{}^{\lambda}\sigma_{\lambda \nu}+u^\lambda \mathcal{D}_\lambda \sigma_{\mu\nu}+ C_{\mu\alpha\nu\beta}u^\alpha u^\beta \right)\\
&\left.+\left(A_1 bd+\tau_\omega\right) u^\lambda \mathcal{D}_\lambda \sigma_{\mu\nu} \right]\\
&+ b^2 (B_1+2 A_3) \mathcal{D}_\mu \mathcal{D}_\nu \sigma^{\mu\nu}+\ldots \\
\end{split}
\end{equation}

We rewrite the above expression in a more useful form by isolating the terms that are manifestly non-negative (keeping terms containing no more than three derivatives):
\begin{equation}
\label{diver}
\begin{split}
4G_{\text{AdS}}b^{d-1} \mathcal{D}_\mu J^\mu_S &= \frac{2b}{d} \left[ \sigma_{\mu\nu} -\frac{bd(d-2)}{2}\left(A_3 -\frac{2 A_2}{d-2} \right) \omega_{\mu\lambda}\omega^\lambda{}_\nu\right.\\
&-\frac{bd(d-2)}{2}\left(A_3 + \frac{1}{d(d-2)}\right)\left(\sigma_{\mu}{}^{\lambda}\sigma_{\lambda \nu}+u^\lambda \mathcal{D}_\lambda \sigma_{\mu\nu}+ C_{\mu\alpha\nu\beta}u^\alpha u^\beta \right)\\
&\left. +\frac{1}{2}\left(A_1 bd+\tau_\omega\right) u^\lambda \mathcal{D}_\lambda \sigma_{\mu\nu} \right]^2\\
&+ b^2 (B_1+2 A_3) \mathcal{D}_\mu \mathcal{D}_\nu \sigma^{\mu\nu}+\ldots \\
\end{split}
\end{equation}
The second law requires that the right hand side of the above equation be positive semi-definite at every point on the boundary. This gives us the single 
constraint :
\begin{equation}
\label{constraint}
B_{1} +2\,A_{3} = 0 
\end{equation}

Equation \eqref{constraint} is the main result of this subsection. Any Weyl covariant 
entropy current that obeys the constraint spelt out in \eqref{constraint} 
has a manifestly non negative divergence of the entropy current, keeping 
only terms to the order of interest.

A particular example of such a current was constructed in \cite{Loganayagam:2008is}. The entropy current proposed in \cite{Loganayagam:2008is} is equivalent to the following proposal for the coefficients
\begin{equation}\label{eq:JSparamSimp}
\begin{split}
\tilde{A}_1 &= -\frac{\tau_\omega}{bd}\ , \quad\ \tilde{A}_2 = -\frac{1}{2d} , \quad\ \tilde{B}_2 = \frac{1}{d-2}\ , \quad\ \tilde{B}_1= -2 \tilde{A}_3 =\frac{2}{d(d-2)}
\end{split}
\end{equation}
which yields a simple manifestly non-negative formula for the rate of entropy production $ T\mathcal{D}_\mu \tilde{J}^\mu_S = 2\eta\ \sigma_{\mu\nu} \sigma^{\mu\nu}+\ldots $.

Another example of an entropy current whose divergence is non-negative
 is the entropy current derived from gravity in the previous section 
using the coefficients appearing in the equation~\eqref{eq:JSparam}. 
Since the values of $A_{3}$ and $B_{1}$ appearing in \eqref{eq:JSparam}
satisfy the constraint \eqref{constraint}, we conclude that the entropy current constructed in the previous subsection satisfies the second law. More explicitly, by substituting the values of $A$ and $B$ coefficients from \eqref{Jsfinal:eq}, we  get the rate of entropy production as
\begin{equation}
\label{entropyprodn}
\begin{split}
 \mathcal{D}_\mu J^\mu_S &= \frac{2\eta}{T} \left[ \sigma_{\mu\nu} 
+\frac{1}{2}\left(A_1 bd+\tau_\omega\right) u^\lambda \mathcal{D}_\lambda \sigma_{\mu\nu} \right]^2+\ldots\\
\end{split}
\end{equation}
where $A_1$ and $\tau_\omega$ have been defined in equations \eqref{Jsfinal:eq} and \eqref{enparam:eq} respectively.

\section{Black Holes in AdS}
\subsection{AdS Kerr metrics as fluid duals}

In the previous sections, we have found the bulk dual to
arbitrary fluid dynamical evolutions on the boundary, 
to second order in the derivative expansion. In this section,
we now proceed to test our results against a class of exact solutions 
of Einstein's equations. 

This class of solutions is the set of rotating black holes in
the global $AdS$ spaces. The dual boundary stress tensor to these 
solutions varies on the length scale unity (if we choose our boundary 
sphere to be of unit radius). On the other hand the temperature of these 
black holes may be taken to be arbitrarily large. It follows that, in the 
large temperature limit, these black holes are dual to `slowly varying' 
field theory configurations that should be well described by fluid dynamics. 
All of these remarks, together with nontrivial evidence for this 
expectation was described in \cite{Bhattacharyya:2007vs}. In this subsection,
we will complete the programme initiated in \cite{Bhattacharyya:2007vs} for
uncharged blackholes by demonstrating that the full bulk metric of these
high temperature rotating black holes agrees in detail with the 2nd order
bulk metric determined by our analysis earlier in this paper. This exercise
was already carried out in \cite{Bhattacharyya:2008ji} for the special case $d=4$. 

Consider the AdS-Kerr BHs in arbitrary dimensions - exact solution for the rotating blackholes in general AdS$_{d+1}$   in different coordinates is derived in reference \cite{Gibbons:2004uw}. Following \cite{Gibbons:2004uw}, we begin by defining two integers $n$ and $\epsilon$ via $d= 2 n+\epsilon$ with $\epsilon= d\ \text{mod}\ 2$. We can then parametrise the $d+1$ dimensional AdS Kerr solution by a radial co-ordinate $r$, a time co-ordinate $\hat{t}$ along with $d-1= 2 n+\epsilon-1$ spheroidal co-ordinates on S$^{d-1}$. We will choose these spheroidal co-ordinates to be $n+\epsilon$ number of direction cosines $\hat{\mu}_i$ (obeying $\sum_{k=1}^{n+\epsilon}\hat{\mu}_k^2 = 1$ ) and $n+\epsilon$ azimuthal angles $\hat{\varphi}_i$ with $\hat{\varphi}_{n+1}=0$ identically. The angular velocities along the different $\hat{\varphi}_i$s are denoted by $a_i$ ($a_{n+1}$is taken to be zero identically).

In this `altered' Boyer-Lindquist co-ordinates, AdS Kerr metric assumes the form (See equation (E.3) of the \cite{Gibbons:2004uw})
\begin{equation}\label{gibbons:eq}
\begin{split}
ds^2 &= - W(1 +r^2) d\hat{t}^2 + \frac{\mathfrak{F} dr^2}{1-2M/V} + \frac{2M}{V\mathfrak{F}}
\left(W d\hat{t} -\sum_{i=1}^{n} \frac{a_i \hat{\mu}_i^2 d\hat{\varphi}_i}{1 - a_i^2}\right)^2 \\
& + \sum_{i=1}^{n+\epsilon} \frac{r^2 + a_i^2}{1 - a_i^2}\left[d\hat{\mu}_i^2+\hat{\mu}_i^2 d\hat{\varphi}_i^2 \right]
 - \frac{1}{W(1+ r^2)}
    \left( \sum_{i=1}^{n+\epsilon} \frac{r^2 + a_i^2}{1- a_i^2}
    \hat{\mu}_i d\hat{\mu}_i\right)^2 
\end{split}
\end{equation}
where
\begin{equation}\label{gibbonsWVF:eq}
\begin{split}
W \equiv \sum_{i=1}^{n+\epsilon} \frac{\hat{\mu}_i^2}{1- a_i^2} \quad;\quad
V \equiv r^{d}\, (1+\frac{1}{r^2})\, \prod_{i=1}^n (1+ \frac{a_i^2}{r^2})\quad\text{and}\quad
\mathfrak{F} \equiv \frac{1}{1+r^2} 
  \sum_{i=1}^{n+\epsilon} \frac{r^2 \hat{\mu}_i^2}{r^2+a_i^2}
\end{split}
\end{equation}

We first perform a co-ordinate transformation of the form
\begin{equation}\label{coordchange:eq}
\begin{split}
d\hat{t} &= dt - \frac{dr}{\left(1 + r^2\right)\left(1 -2M/V\right)}  \quad;\quad
d\hat{\varphi}_i = d\varphi_i - \frac{a_i dr}{\left(r^2 + a_i^2\right)\left(1 -2M/V\right)}
\end{split}
\end{equation}
followed by another transformation of the form
\begin{equation}\label{muiredef:eq}
\begin{split}
\mu_i^2 &\equiv \frac{1}{W}\left(\frac{\hat{\mu}_i^2}{1-a_i^2}\right) \quad \text{with} \quad W = \frac{1}{1 - \sum_i a_i^2\mu_i^2} \quad,\quad \mathfrak{F}=W\left[\sum_i \frac{
\mu_i^2}{1+\frac{a_i^2}{r^2}}-\frac{
1}{1+\frac{1}{r^2}}\right]
\end{split}
\end{equation}
to get
\begin{equation}\label{kerrmet1}
\begin{split}
ds^2 &= -2 u_\mu dx^\mu(dr+r\mathcal{A}_\nu dx^\nu) + \left[r^2 g_{\mu\nu}+\Sigma_{\mu\nu}\right] dx^\mu dx^\nu +\frac{u_\mu u_\nu}{V \mathfrak{F} b^d}  dx^\mu dx^\nu 
\end{split}
\end{equation}
where 
\begin{equation}\label{kerrmet2}
\begin{split}
u^\mu \partial_\mu &\equiv \partial_{t}+a_i \partial_{\varphi_i}\quad,\quad \mathcal{A}_{\mu} = 0  \quad,\quad b\equiv(2M)^{-1/d}\\
g_{\mu\nu} &\equiv W\left[-dt^2 + \sum_i\left(d\mu_i^2 + \mu_i^2 d\varphi_i^2\right)\right]\\
\Sigma_{\mu\nu}&\equiv W\left[-dt^2 +\sum_i a_i^2\left(d\mu_i^2 + \mu_i^2 d\varphi_i^2\right) + \left(\sum_ia_i^2\mu_i d\mu_i\right)^2\right]
\end{split}
\end{equation}

This expression can be further simplified using the following identities

\begin{equation}\label{kerrmet3}
\begin{split}
\Sigma_{\mu\nu}&= u_{(\mu}\mathcal{S}_{\nu)\lambda}u^\lambda -\omega_{\mu}{}^{\lambda}\omega_{\lambda\nu}\\
r^2 V \mathfrak{F} &= \text{det}\left[r\ \delta^\mu_{\nu}-\omega^\mu{}_{\nu}\right]
\end{split}
\end{equation}
where the determinant of a tensor $M^{\lambda}_\sigma$ is defined by
\[ \epsilon_{\mu\nu\ldots}M^{\mu}_\alpha M^{\nu}_\beta \ldots = \text{det}\left[M^{\lambda}_\sigma \right] \epsilon_{\alpha\beta\ldots} \]

Hence, we conclude that the AdS Kerr metric in arbitrary dimensions can be rewritten in the form
\begin{equation}\label{kerrmet}
\begin{split}
ds^2&=-2 u_\mu dx^\mu \left( dr + r\ \mathcal{A}_\nu dx^\nu \right) + \left[ r^2 g_{\mu\nu} +u_{(\mu}\mathcal{S}_{\nu)\lambda}u^\lambda -\omega_{\mu}{}^{\lambda}\omega_{\lambda\nu}\right]dx^\mu dx^\nu\\ 
&\qquad+ \frac{r^2 u_\mu u_\nu}{b^d\text{det}\left[r\ \delta^\mu_{\nu}-\omega^\mu{}_{\nu}\right]}  dx^\mu dx^\nu\\ 
\end{split}
\end{equation}
We have checked this form explicitly using Mathematica till $d=8$.

This metric can also be written in the form
\begin{equation}
\begin{split}
&ds^2=-2 u_\mu dx^\mu \left( dr + r\ \mathcal{A}_\nu dx^\nu \right) + r^2 g_{\mu\nu} dx^\mu dx^\nu\\
&-\left[\omega_{\mu}{}^{\lambda}\omega_{\lambda\nu}+\frac{1}{d-2}\mathcal{D}_\lambda \omega^{\lambda}{}_{(\mu}u_{\nu)}+\frac{1}{(d-1)(d-2)}\mathcal{R}u_{\mu}u_{\nu}\right]dx^\mu dx^\nu \\
&+ \frac{r^2 u_\mu u_\nu}{b^d\text{det}\left[r\ \delta^\mu_{\nu}-\omega^\mu{}_{\nu}\right]}  dx^\mu dx^\nu\\ 
\end{split}
\end{equation}
or alternatively
\begin{equation}\label{eq:VGexpAdSKerr} 
\begin{split}
ds^2 &= -2 u_\mu dx^\mu(dr+\mathcal{V}_\nu dx^\nu)+\mathfrak{G}_{\mu\nu}dx^\mu dx^\nu \qquad \text{with}\\
\mathcal{V}_\mu &= r \mathcal{A}_\mu -\mathcal{S}_{\mu\lambda}u^\lambda- \frac{r^2 u_\mu }{2b^d\text{det}\left[r\ \delta^\mu_{\nu}-\omega^\mu{}_{\nu}\right]} \\
\end{split}
\end{equation}
\begin{equation*}
\begin{split}
\mathfrak{G}_{\mu\nu}&= r^2 P_{\mu\nu}-\omega_{\mu}{}^{\lambda}\omega_{\lambda\nu}
\end{split}
\end{equation*}

It is easily checked that this metric agrees(upto second order in boundary derivative expansion) with the metric presented in \eqref{GSexp:eq} in
 section 4 of this paper, upon inserting the velocity and temperature fields 
listed in \eqref{kerrmet2}.

%_____________________________________________
\subsection{The Energy momentum tensor and the Entropy Current for the AdS Kerr Black Hole}
\label{blackentropy}
%_____________________________________________

The exact energy momentum tensor for the AdS Kerr Black Hole described can be computed
using the standard counterterm methods. The non-anomalous part of the energy momentum tensor is given by 
\begin{equation}\label{enmomBH:eq}
\begin{split}
T_{\mu\nu} = p(g_{\mu\nu}+du_\mu u_\nu)\quad \text{with}\quad p=\frac{1}{16\pi G_{AdS}b^d}
\end{split}
\end{equation}
which is consistent with \eqref{enparam:eq} if we take into account the fact that $\sigma_{\mu\nu}=0$ in these configurations. 

The equation for the event horizon of the AdS Kerr Black Hole is given by $V=2M$ or 
\begin{equation}\label{horizon:eq}
\begin{split}
\frac{1}{b^d} &= \rH^{d}\, (1+\frac{1}{\rH^2})\, \prod_{i=1}^n (1+ \frac{a_i^2}{\rH^2})\\
&= \rH^{d}\left[1+\frac{1+\sum_i a_i^2}{\rH^2}+\ldots \right]
\end{split}
\end{equation}
which can be solved for $\rH$ to give
\begin{equation}
\begin{split}
\rH &= \frac{1}{b}\left[1-\frac{b^2}{d}\left(1+\sum_i a_i^2\right)+\ldots \right]\\ &=\frac{1}{b}-\dfrac{d+2}{2d(d-2)}b\ \omega^{\mu \nu}\omega_{\mu\nu}- \frac{b\ \mathcal{R}}{d(d-2)(d-1)} +\ldots
\end{split}
\end{equation}
which agrees with the expression for the event horizon in \eqref{eq:rHfinal} 
upon inserting the velocity and temperature field configurations 
\eqref{kerrmet2}.

The entropy current for the AdS Kerr blackhole can be directly obtained from \eqref{Jseq}. We have the following exact results :
\begin{equation}\label{exactJs:eq}
\begin{split}
\sqrt{\text{det}^{(n)}_{d-1}\mathcal{H}}
&= \rH^{d-1}\prod_{i=1}^{n}(1+\frac{a_{i}^{2}}{\rH^2})
= \frac{\rH}{b^d(\rH^2+1)} \\
n^\mu\partial_\mu &= \partial_t +\sum_i\frac{\rH^2+1}{\rH^2+a_{i}^{2}}a_i\partial_{\varphi_i}= u^\mu\partial_\mu +\sum_i\frac{1-a_{i}^{2}}{\rH^2+a_{i}^{2}}a_i\partial_{\varphi_i} \\
J^\mu_S\partial_\mu &= \frac{\rH}{4G_{\text{AdS}}b^d(\rH^2+1)}\left[u^\mu\partial_\mu +\sum_i\frac{1-a_{i}^{2}}{\rH^2+a_{i}^{2}}a_i\partial_{\varphi_i} \right]
\end{split}
\end{equation}

These exact results can alternatively be expanded in a derivative expansion. Keeping terms only upto second order in the derivative expansion, we get
\begin{equation}\label{Gnexp:eq}
\begin{split}
\sqrt{\text{det}^{(n)}_{d-1}\mathcal{H}}
&= \frac{1}{b^{d-1}}[1-\frac{d-1}{d}b^{2}+\frac{b^{2}}{d}\sum_{i}a_{i}^{2}+\ldots]\\
n^\mu\partial_\mu &=  u^\mu\partial_\mu +b^2\sum_i(1-a_i^2)a_i\partial_{\varphi_i}+\ldots \\
\end{split}
\end{equation}
which gives
\begin{equation}
\begin{split}
4G_{\text{AdS}}b^{d-1} J^{\mu}_{S} &= u^{\mu}\left[1-\frac{b^2}{2d}\omega^{\alpha \beta} \omega_{\alpha \beta}-\frac{b^2\mathcal{R}}{d(d-2)}\right]
+\frac{b^2}{d-2}\mathcal{D}_{\lambda} \omega^{\mu \lambda} \\
\end{split}
\end{equation}
We have checked this form explicitly using Mathematica till $d=8$.
 
% so that
% \begin{equation}
% 4\, G_{N}^{d+1}\, b^{d\, -\, 1}\, J^{\mu}_{S} = [\,\prod_{i=1}^{N}\,(\,r_{H}^{2}\,+\,a_{i}^{2}\,)\,]\,n^{\mu}
% \end{equation}
% 
% Expanding upto second order, from $\eqref{h}$ and $ \eqref{n} $, we get :
% \begin{equation}
% \label{hf}
% \sqrt{h} = \dfrac{1}{b^{d\,-\,1}}\,[\,1\,-\,\dfrac{d\,-\,1}{d}\,b^{2}\,+\,\dfrac{b^{2}}{d}\,\sum_{i=1}^{N}\,a_{i}^{2}\,]
% \end{equation}
% \\
% and
% \begin{equation}
% (\delta n)^{t} = 0 \hspace{9 mm}(\delta n)^{\mu_{i}} = 0\hspace{9 mm}(\delta n)^{\phi_{i}} = \dfrac{a_{i}(1\,-\,a_{i}^{2})}{r_{0}^{2}}
% \end{equation}
% \\
% ( where $ n^{\mu}\,=\,u^{\mu} \,+ \,\delta n^{\mu} $  upto second order ), so that
% \begin{equation}
% \begin{split}
% 4\, G_{N}^{d+1}\, b^{d\, -\, 1}\, J^{\mu}_{S} &= u^{\mu}\,[\,1\,-\,\dfrac{b^{2}}{2\,d}\,\omega^{\alpha \beta}\, \omega_{\alpha \beta} \,+\,\dfrac{\mathcal{R}}{d\, (d\, -\, 2)}\,] \\
% &\hspace{6 mm} +\,\dfrac{b^{2}}{d\, -\, 2}\, \mathcal{D}_{\lambda}\, \omega^{\mu \lambda} \\
% \end{split}
% \end{equation}
% \\
% \textit{(found with the help of the computation done on Mathematica)} .
% \\
Comparing the above with \eqref{Jsfinal:eq} and remembering that $\sigma^{\alpha \beta}=0$ for the AdS Kerr black hole, we find that our results in the previous sections are consistent with these exact solutions.

\section{Discussion}

In this paper we have constructed an explicit map from solutions of the
(generalized) Navier Stokes equations on a $d-1,1$ dimensional boundary 
with an arbitrary weakly curved metric $g_{\mu \nu}$ to the space of 
regular solutions to the Einstein equations with a negative cosmological
constant that asymptote, at small $z$, to $ds^2 =z^{-2}\left[dz^2 +
g_{\mu\nu} dx^\mu dx^\nu\right]$. We have demonstrated that our solution 
space is exhaustive locally in solution space. In other words consider 
a particular bulk solution $B$ that is dual, under the map constructed 
in this paper, to a fluid flow $F$. Then every regular slowly varying 
bulk solution to Einstein's equations that is infinitesimally separated 
from $B$ is dual to a fluid flow infinitesimally separated from $F$. 

We have also demonstrated that - subject to certain restrictions on the 
long time behavior  - all the metrics constructed in this paper have 
regular event horizons, and have constructed the event horizon manifold 
of our solutions in this paper. It would be interesting to relax the 
restrictions on long time behavior under which this result follows, and 
simultaneously examine under what conditions these restrictions are 
dynamically automatic from the equations of fluid dynamics. In particular, 
as the long time limit of a fluid flow on a static metric is necessarily 
non dissipative, it would be interesting to fully classify all nondissipative
flows on static background geometries. \footnote{It is natural to guess 
that this set is exhausted by uniform motion (in the case of the boundary
R$^{d-1,1}$) and rigid rotations (in the case of the boundary S$^{d-1,1}$), but we 
are unaware of proofs if any of this intuition. We thank G.Gibbons for discussions on this issue.}

We have been able to put our construction of the event horizons described 
above to practical use: by pulling the area form on the event horizon back 
to the boundary, we have been able to define an entropy current for the 
dual fluid flow. The divergence of this current is guaranteed to be non
negative by the classic area increase theorem of black hole physics. The 
entropy current we have constructed is a sort of local `Boltzmann H 
function' which can, locally, only be created and never destroyed. The
local entropy increase theorem establishes the irreversible nature of 
dual fluid flows. It may be interesting to study the structure of gradient 
flows generated by this `entropy function'.

In this paper we have presented explicit expressions, to second order in
the derivative expansion, for the holographic stress tensor of boundary
fluid dynamics. Our expression  in equation~\eqref{enmom:eq}, which is only a slight
extension of the expression listed in \cite{Haack:2008cp}, is extremely simple.
It is clear from a glance at equation~\eqref{enmom:eq} that several ratios of the 
transport and thermodynamical coefficients listed there are `universal', i.e. 
independent of the spacetime dimension for conformal fluids with a gravitational 
dual. It would be very interesting to investigate whether any of this universality persists for more general (e.g. charged or non conformal) fluids with a gravitational dual. 
It would be useful to have more data in hand before speculating further; we hope to 
return to this issue in the future. 

It would also be useful to understand the second order transport coefficients 
listed in our paper more physically (perhaps in terms of an effective model 
of microscopic dynamics that replaces kinetic theory). Along another line, 
it should not be difficult to derive expressions for these coefficients 
in terms of two and three point correlations functions of the stress tensor. 
In particular  expect the expression for the coefficients of the nonlinear 
terms in equation~\eqref{enmom:eq} to be determined from the three point LSZ-type 
formula (obtained by stripping this correlator of its hydrodynamic poles 
and extracting the coefficient of $k^2$ in the residue). 

It would be interesting to attempt to semiclassically quantize the 
space of hydrodynamic (and related) solutions of gravity with a negative 
cosmological constant. Such an exercise could begin to allow one to account 
for the effects of statistical fluctuations about hydrodynamical flows. 

Finally, the map developed in this paper should allow us to incorporate 
atleast a small fraction of the enormous (180 year old) study of 
hydrodynamics into the study of gravity. We hope to return to this issue
in the future. 

\subsection*{Acknowledgements}

We would like to acknowledge useful discussions and correspondences
with J.Bhattacharya,  M.Haack,  V.Hubeny,  G.Mandal, T.Morita,  M.Rangamani,
H. Reall,  S.Trivedi,  S.Wadia  and L.G.Yaffe. Further, We would like to 
acknowledge several useful discussions with and participants of 
the ICTS Monsoon workshop on string theory held in TIFR, Mumbai. 
We would also like to acknowledge useful discussion with the 
students in the TIFR theory room. Several of the calculations 
of this paper were performed using the excellent Mathematica
package \texttt{diffgeo.m} developed by Matthew Headrick.

The work of S.M. was supported in part by Swarnajayanti Fellowship. I.M.
would like to acknowledge the hospitality of TIFR during 
the monsoon workshop, when this work was done. A.S. would like
to acknowledge the Visting Student Research Programme(VSRP) at
TIFR for support while this work was conducted. We would all like to 
acknowledge our debt to the people of India for their generous
and steady support to research in the basic sciences.

%%%%%%%%%%%%%%%%%%%%%%%%%%%%%%%%%%%%%%%%%%%%%%%%%%%%%%%%%%%%%%%%%%%%%%%%%%
\section*{Appendices}
\appendix
%%%%%%%%%%%%%%%%%%%%%%%%%%%%%%%%%%%%%%%%%%%%%%%%%%%%%%%%%%%%%%%%%%%%%%%%%%

\section{Notation}\label{app:notation}

We work in the $(-++\ldots)$ signature. The dimensions of the spacetime in which the conformal fluid lives is denoted by $d$ . We usually assume $d>2$ unless otherwise specified. In the context of AdS/CFT, the dual AdS$_{d+1}$ space has $d+1$ spacetime dimensions.The greek indices $\mu,\nu= 0,1,\ldots,d-1$ are used as boundary space-time indices, whereas the latin indices $A,B=0,1,\ldots, d$ are used as the bulk indices. Throughout this paper, we take the extra holographic co-ordinate $y^{(d)}=r$ with the boundary of the bulk spacetime at $r\rightarrow\infty$.

We take the bulk AdS radius to be unity which is equivalent to setting the bulk cosmological constant to be $\Lambda_{\text{AdS}}=-\frac{d(d-1)}{2}$. We denote the bulk Newton constant by $G_{\text{AdS}}$. For the ease of reference, we now give the value of $G_{\text{AdS}}$ for some of the well-known CFTs with gravity duals : (See \cite{Maldacena:1997re,Aharony:1999ti,Aharony:2008ug} for further details)
\begin{enumerate}
\item The d=4, $\mathcal{N}$=4 Super Yang-Mills theory on $N_c$ D3-Branes with a gauge group SU(N$_c$) and a `t Hooft coupling $\lambda\equiv g_{YM}^2 N_c$ is believed to be dual to IIB string theory on  AdS$_5\times$S$^5_{R=1}$ with  $G_{\text{AdS}_5}=\pi /(2 N_c^2)$ and $\alpha'=(4\pi\lambda)^{-1/2} $ .
\item A d=3, $\mathcal{N}$=6 Superconformal \footnote{In the case of $k=1,2$, the supersymmetry should get enhanced to d=3, $\mathcal{N}$=8.} Chern-Simons theory on $N_c$ M2-Branes with a gauge group U(N$_c$)$_k\times$ U(N$_c$)$_{-k}$ (where the subscripts denote the Chern-Simons couplings) and a `t Hooft coupling $\lambda\equiv N_c/k$ is conjectured  to be dual to M-theory on AdS$_4\times$S$^7_{R=2}$/Z$_k$ with $G_{\text{AdS}_4}=N_c^{-2}\sqrt{9\lambda/8}=3k^{-1/2}(2N_c)^{-3/2}$.
\item A d=6 superconformal theory on  $N_c$ M5-Branes is conjectured  to be dual to M-theory on AdS$_7\times$S$^4_{R=1/2}$ with $G_{\text{AdS}_7}=3\pi^2/(16 N_c^{3})$.
\end{enumerate}
These values can be used easily to compare our results with the standard results in the literature.

We use round brackets to denote symmetrisation and square brackets to denote antisymmetrisation. For example, $B_{(\mu\nu)}\equiv B_{\mu\nu}+B_{\nu\mu}$ and $B_{[\mu\nu]}\equiv B_{\mu\nu}-B_{\nu\mu}$. Our conventions for Christoffel symbols and the curvature tensors are fixed by the relations
\begin{equation}
\begin{split}
\nabla_{\mu}V^{\nu}&=\partial_{\mu}V^{\nu}+\Gamma_{\mu\lambda}{}^{\nu}V^{\lambda} \qquad \text{and}\qquad [\nabla_\mu,\nabla_\nu]V^\lambda=-R_{\mu\nu\sigma}{}^{\lambda}V^\sigma .
\end{split}
\end{equation}

We have included a table with other useful parameters used in the text. In the table~\ref{notation:tab}, the relevant equations are denoted by their respective equation numbers appearing inside parentheses.

\begin{table}\label{notation:tab}
 \centering
 \begin{tabular}{||r|l||r|l||}
   \hline
   \multicolumn{4}{||c||}{\textbf{Table of Notation}} \\
   \hline 
   Symbol & Definition & Symbol & Definition \\
   \hline
   $u^\mu$ & Fluid velocity \eqref{eqn:branest} & $T$ & Fluid temperature\\
   $g_{\mu\nu}$ & Boundary metric & $P_{\mu\nu}$ & $g_{\mu\nu}+u_\mu u_\nu$ \\
   $b$ & $d/(4\pi T)$ & $p$ & Fluid pressure \eqref{enmom:eq}\\
   $\mathcal{A}_\mu$ & Weyl-Connection \eqref{eqn:Adef} & $\mathcal{D}_\mu$ & Weyl-Covariant  \\
   & & & derivative \eqref{eqn:Ddef}\\
   $\mathcal{R}_{\mu\nu\lambda\sigma}$ & Weyl-covariantized  &$\mathcal{S}_{\mu\nu}$ & Weyl-covariantized \\
    & Riemann curvature \eqref{eqn:Rdef} & & Schouten tensor \eqref{eqn:Sdef}\\
   $\mathcal{F}_{\mu\nu}$ & $\nabla_{[\mu}\mathcal{A}_{\nu]}$ (see \eqref{eqn:Rdef}) & $u^\lambda\mathcal{D}_{\lambda}\sigma_{\mu\nu}$ & See \eqref{eqn:sigmaRdef}
   \\
   $\mathcal{R}_{\mu\nu},\mathcal{R}$ & Weyl covariantized  & $C_{\mu\nu\lambda\sigma}$ & Boundary Weyl  \\
    & Ricci tensor/scalar \eqref{eqn:sigmaRdef} & & curvature tensor \eqref{eqn:Cdef} \\
   \hline
   $\sigma_{\mu\nu}$ & Shear strain rate, $1/2\ \mathcal{D}_{(\mu}u_{\nu)}$ & $\omega_{\mu\nu}$ & Fluid vorticity, $1/2\ \mathcal{D}_{[\mu}u_{\nu]}$ \\
   $\eta$ & Shear Viscosity \eqref{enmom:eq} & $\tau_{\omega}$ & Shear relaxation time \eqref{enmom:eq} \\
   $\tau_{1,2}$ & Second order transport  & $\tau_{\Pi},\kappa$& Alternate notation for \\
   $\xi_{\sigma,\omega,C}$& \quad coefficients \eqref{Tgeneral:eq}&$\lambda_{1,2,3}$& \quad 2nd order coefficients \eqref{baierrelation:eq} \\
  \hline
   $G_{AB}$ & AlAdS$_{d+1}$ Bulk metric & $\mathcal{G}_{AB}$ & Bulk Einstein tensor \\
   $\mathcal{V}_\mu,\mathfrak{G}_{\mu\nu}$ & Defined by \eqref{eq:metstd} & $(\mathfrak{G}^{-1})^{\mu\nu}$ & Defined by \eqref{eq:invmetstd}\\
   $\xi^A$ & Horizon Normal vector \eqref{eq:normstd} & $n^\mu,\kappa^\mu$  & See \eqref{eq:normstd}\\
   $H_{\mu\nu}$ & Induced metric  & $\rH(x)$ & Position of the horizon\\
 & on Horizon \eqref{eq:HorInd} and \eqref{eq:Hdef} & & given by $r=\rH(x)$ \\
   $h_i$ & Coefficients in \eqref{eq:rHfinal} & $A_i,B_i$ & See \eqref{Jsfinal:eq} \\
   $F,H_i,K_i,L$ & See \eqref{metricfns:eq} & &\\
  \hline
\end{tabular}
\end{table}

\section{d=2}

Through the text of this paper we have worked with conformal fluids in $d>2$ 
dimensions. In this section we explain that conformal fluid dynamics in $d=2$ 
is special and essentially trivial. 

To start with note that a traceless stress tensor in 
$d$ dimensions has $s_d=\frac{d^2+d-2}{2}$ independent components. The 
assumption of local thermalization in the fluid dynamical limit allows us to 
work instead with the $d$ variables of fluid dynamics; 
the velocities and temperature. Now $d<s_d$ for $d>2$; it is precisely for 
this reason that fluid dynamics contains physical information beyond the 
conservation of the stress tensor.  However $s_2=2$;  consequently 
two dimensional conformal fluid dynamics is simply the assertion of 
conservation of the two dimensional stress tensor. One may as well 
work directly with the components of the stress tensor. The general 
solution to the conservation of the stress tensor in $d=2$ is of course 
well known. In a frame in which 
the boundary metric locally takes the form 
$ds^2=e^{2 \phi} dx^+ dx^-$ (and ignoring anomaly effects in this discussion) 
the most general conserved and traceless stress tensor is given by $T_{++}=
f(x^+)$ and $T_{--}=g(x^-)$ for arbitrary functions 
$f$ and $g$. This constitutes the most general solution to `conformal fluid 
dynamics' in two dimensions. Note that according to this solution, left and 
right moving waves do not interact with each other. Consequently two dimensional conformal `fluid' dynamics is both trivial and a misnomer; conformal 
fluids in two dimensions do not locally equilibrate.

The triviality of conformal fluid dynamics in two dimensions has a simple 
gravitational counterpart: every solution of Einstein's equations in two 
dimensions is locally $AdS_3$. All generally coordinate inequivalent regular 
solutions of these equations are the BTZ black holes. (Note that the point 
mass solutions, studied extensively for instance in \cite{David:1999zb}, 
have a naked singularity atleast from the purely gravitational point of view). 
Conformally inequivalent slicings of the same geometry (a la Brown and Hanneaux)
generate the left and right moving waves described in the previous subsection. From the 
bulk point of view these solutions are trivial because they are all 
(large) diffeomorphism equivalent to static black holes. 

There is yet another way to express the triviality of conformal fluid 
dynamics in two dimensions. It turns out that there are no non-zero 
Weyl-covariant quantities which can be formed out of velocity/temperature derivatives and hence, as noted by \cite{Kajantie:2007bn,Haack:2008cp}, the first order fluid dynamical metric becomes an exact solution of the bulk Einstein equations (see section 4 of \cite{Haack:2008cp} for more details). For all the reasons 
spelt out above, in the rest of our paper we will focus on $d>2$.

\section{Pointwise solution to dynamics at second order in derivatives}

As explained in \cite{Bhattacharyya:2008jc}, in order to construct the 
map from solutions of fluid dynamics to solutions of gravity at second 
order, we need to `solve' the equations of fluid dynamics, at a point 
$x^\mu$ to second order in derivatives. While it is of course very difficult 
to find the general global solutions to fluid dynamics, the corresponding 
equations are very easily solved at a point. In this Appendix  
we review the solution of these equations in explicitly Weyl covariant terms. 
The results of this appendix were utilized in our construction of the bulk 
metric in section 4. 

For solving the bulk constraint equations upto second order, we need ($\mathcal{D}_{\mu}T^{\mu \nu}=0$) evaluated upto second order 
\begin{equation}\label{DTcons:eq}
\mathcal{D}_{\mu}b = 2 b^2 \frac{4\pi\eta}{s}\left[ \frac{\sigma_{\alpha\beta}\sigma^{\alpha \beta}}{d-1} u_{\mu} - \frac{\mathcal{D}_{\lambda}{\sigma^{\lambda}}_{\mu}}{d}  \right]+\ldots\\
\end{equation}
where we have introduced the entropy density $s$ of the conformal fluid related to its pressure by $s=pd/T=4\pi pb$. This can be used to solve for the partial derivatives of $b$ completely in terms of velocity derivatives  
\begin{equation}\label{derivb:eq}
\begin{split}
\partial_{\mu}b &= \mathcal{A}_{\mu} b +2 b^2 \frac{4\pi\eta}{s}\left[ \frac{\sigma_{\alpha\beta}\sigma^{\alpha \beta}}{d-1} u_{\mu} - \frac{\mathcal{D}_{\lambda}{\sigma^{\lambda}}_{\mu}}{d}  \right]+\ldots\\
\partial_{\mu}\partial_{\nu}b&=b(\partial_{\mu}\mathcal{A}_{\nu} + \mathcal{A}_{\mu}\mathcal{A}_{\nu})+\ldots
\end{split}
\end{equation}
Since the left hand side of the last equation is symmetric in $\mu$ and $\nu$, we get an integrability condition
\begin{equation}\label{integb:eq}
\partial_{\mu}\mathcal{A}_{\nu}=\partial_{\nu}\mathcal{A}_{\mu}+\ldots
\end{equation}

Hence, we conclude that to this order we have a valid fluid configuration 
in a patch around a point $P_0$ provided we assume
\begin{equation}\label{bsoln:eq}
\begin{split}
b&=b_{0} + \epsilon b_{0} \mathcal{A}_{\nu0} x^{\nu} +2 \epsilon^2 b_{0}^2 \frac{4\pi\eta}{s}\left[ \frac{\sigma_{\alpha\beta}\sigma^{\alpha \beta}}{d-1} u_{\mu} - \frac{\mathcal{D}_{\lambda}{\sigma^{\lambda}}_{\mu}}{d}  \right]_0\\
&+ \epsilon^{2} \frac{b_{0}}{2} \left[\partial_{\mu}\mathcal{A}_{\nu} + \mathcal{A}_{\mu}\mathcal{A}_{\nu}\right]_0 x^{\mu}x^{\nu} + \ldots\\
F_{\mu\nu}&\equiv \partial_{[\mu}\mathcal{A}_{\nu]} = 0+\ldots\\
\end{split}
\end{equation}
For the metric given in the text to be a solution of the Einstein equations, it is necessary that the velocity/temperature fields obey the above equations of motion with $\eta/s=1/(4\pi)$. 

\nocite{Policastro:2001yc,
Son:2002sd,Policastro:2002se,Herzog:2002fn,Policastro:2002tn,Herzog:2002pc,
Herzog:2003ke,Kovtun:2003vj,Buchel:2003ah,Kovtun:2003wp,Buchel:2003tz,
Kovtun:2004de,Buchel:2004hw,Buchel:2004di,Buchel:2004qq,
Kovtun:2005ev,Starinets:2005cy,Aharony:2005bm,Janik:2005zt,
Mas:2006dy,Son:2006em,Saremi:2006ep,Maeda:2006by,CasalderreySolana:2006rq,
Gubser:2006bz,Janik:2006gp,Liu:2006nn,Nakamura:2006ih,Janik:2006ft,Benincasa:2006fu,
Lin:2006rf,Saremi:2007dn,Heller:2007qt,Son:2007vk,Kovchegov:2007pq,Lahiri:2007ae,
Myers:2007we,Bhattacharyya:2007vs,Lin:2007fa,Chesler:2007sv,Ejaz:2007hg,Kats:2007mq,
Benincasa:2007tp,Baier:2007ix,Bhattacharyya:2008jc,Natsuume:2007ty,Natsuume:2007tz,
Kajantie:2008rx,Natsuume:2008iy,Loganayagam:2008is,Buchel:2008ac,VanRaamsdonk:2008fp,
Choi:2008he,Gubser:2008vz,Bhattacharyya:2008xc,Chernicoff:2008sa,
Buchel:2008xr,Myers:2008fv,Dutta:2008gf,Siopsis:2008xz,Gubser:2008pc,Heller:2008mb,Bhattacharyya:2008ji,Buchel:2008bz,Myers:2008yi,Starinets:2008fb,Herzog:2008wg,Haack:2008cp,Natsuume:2008ha,Natsuume:2008gy,Kinoshita:2008dq,Buchel:2008kd,Buchel:2008ae,Miranda:2008vb,Erdmenger:2008rm,Banerjee:2008th}
\bibliographystyle{JHEP}
\bibliography{generalD}

\end{document}